\documentclass[12pt]{iopart}
\usepackage{graphicx,dcolumn,bm,amssymb,xcolor,hyperref}

\begin{document}

\title[Ponderomotive electron scattering off a Gaussian laser focus]{Ultra-intense laser pulse characterization using ponderomotive electron scattering}

\author[1,*]{Felix Mackenroth$^1$, Amol R. Holkundkar$^{1,2}$, Hans-Peter Schlenvoigt$^3$ }
\address{$^1$ Max Planck Institute for the Physics of Complex Systems, N\"othnitzer Strasse 38, 01187 Dresden, Germany}
\address{$^2$ Department of Physics, Birla Institute of Technology and Science - Pilani, Rajasthan, 333031, India}
\address{$^3$ Helmholtz-Zentrum Dresden -- Rossendorf, Institute of Radiation Physics, Bautzner Landstr. 400, 01328 Dresden, Germany}

\ead{mafelix@pks.mpg.de}

\date{\today}

\begin{abstract}
We present a new analytical solution for the equation of motion of relativistic electrons in the focus of a high-intensity laser pulse. We approximate the electron's transverse dynamics in the averaged field of a long laser pulse focused to a Gaussian transverse profile. The resultant ponderomotive scattering is found to feature an upper boundary of the electrons' scattering angles, depending on the laser parameters and the electrons' initial state of motion. In particular, we demonstrate the angles into which the electrons are scattered by the laser scale as a simple relation of their initial energy to the laser's amplitude. We find two regimes to be distinguished in which either the laser's focusing or peak power are the main drivers of ponderomotive scattering. Based on this result, we demonstrate how the intensity of a laser pulse can be determined from a ring-shaped pattern in the spatial distribution of a high-energy electron beam scattered from the laser. We confirm our analysis by means of detailed relativistic test particle simulations of the electrons' averaged ponderomotive dynamics in the full electromagnetic fields of the focused laser pulse.
\end{abstract}

\maketitle

\section{\label{sec:introduction}Introduction}
Recent technological advances in ultra-intense laser systems facilitate studies of particle dynamics in electromagnetic fields of unprecedented strength \cite{VulcanLaser,Mourou_etal_2006,Hooker_etal_2008,Leemans_etal_2010,Zou_etal_2015,Kawanaka_etal_2016,Gales_etal_2018,CORELS, Draco}. In particular, the dynamics of electrons in such ultra-strong fields has been an area of intense research over the past decades \cite{DiPiazza_etal_2012}. It was found, however, that except for specific, highly symmetric cases like plane waves \cite{Sarachik_Schappert_1970,Salamin_Faisal_1996}, the full electron dynamics cannot be solved in closed analytical expressions. In contrast to these computational challenges, however, fully plane laser waves, on the other hand, cannot be realized experimentally, but a laser always has a finite size, overlaying the transverse oscillations driven by the laser's sub-cycle electromagnetic fields by complex envelope dynamics. Over long interaction times, however, the transverse laser field oscillations, are averaged out, leading, e.g., to an exact cancellation of the electrons' energy gain in plane waves according to the Lawson-Woodward theorem \cite{Woodward_1946,Lawson_1979}. Hence, in an interaction that is significantly longer than the laser period it is often sufficient to only consider the envelope's effect on the electrons quivering in the laser field. Such, so-called \textit{ponderomotive scattering} of the electrons was subject of numerous previous publications \cite{LandauI,Schmidt_1979,Boot_etal_1958,Gaponov_Miller_1958,Kibble_1966,Hopf_etal_1976}. In the relativistic regime, a generalized ponderomotive force equation was derived from a Hamiltonian approach \cite{Bauer_etal_1995} with many alternative derivations \cite{Hartemann_etal_1995,Mora_Antonsen_1997,Salamin_Faisal_1997,Hartemann_etal_1998,Quesnel_Mora_1998,Bituk_Fedorov_1999,Narozhny_Fofanov_2000,Yu_etal_2000b,Castillo_Milantev_2014,Shiryaev_2019} and experimental confirmation \cite{Moore_etal_1995,Malka_etal_1997} following. These studies were later used in a series of technical and fundamental applications \cite{Salamin_Mocken_Keitel_2002,Liu_etal_2008}.

Further applications, which are envisaged to be facilitated by a detailed understanding of electron dynamics in realistic laser fields, are reliable metrology schemes for the laser field.  In particular, within the topic of ultra-intense laser physics, direct determination of peak intensity 
is, to date, an unsolved challenge. The typical route for providing this key laser parameter is the combination of three different and distinct measurements: a) pulse energy of the fully amplified, collimated beam; b) pulse duration of a fraction of the collimated beam, typically not at full amplification; and c) focus imaging of typically a not fully amplified and with attenuating elements transported beam. There is a number of shortcomings of this approach: a) does not necessarily yield the energy concentrated within the focus; b) may differ across the beam profile due to radial dispersion and further nonlinear effects \cite{Zhu_etal_2018,Bor_1988}. c) is a time- and spectrum-integrated measurement. Due to the previous effects as well as radial group delay and chromatic aberrations \cite{Heuck_etal_2006,Wu_etal_2018}, the pulse duration in the focus can be much longer than measured with b) and exhibit time structures varying with the focal position. Hence, the typical procedure yields rather an upper limit of peak intensity.

Atomic effects have been proposed and employed to yield a measure of peak intensity \cite{Link_etal_2006,Smeenk_etal_2011,Ciappina_etal_2019}, but were restricted to non-relativistic intensities. At higher intensity, atomic ionization is followed by significant electrodynamic acceleration which can also provide information on peak intensity and focus size \cite{Galkin_etal_2010, Kalashnikov_etal_2015}. That was implemented at mildly relativistic intensities \cite{Ivanov_etal_2018}. Relativistic, intensity-dependent plasma effects like ion acceleration \cite{Fuchs2006NJP, Daido_etal_2012, Macchi_etal_2013,Mackenroth_etal_2017a}, on the other hand, are not feasible since they are quite sensitive to the temporal contrast of the laser pulse \cite{Zeil2010NJP, Draco}, as the plasma formation prior to the pulse peak strongly affects the energy conversion during the peak. Hence, measurements of laser-scattered electron distributions could be a reasonable alternative, especially when relying on beam profile measurements which are much simpler than spectral measurements.

Most of the proposed schemes, however, made strongly simplified assumptions such as modeling the laser as a plane wave \cite{Mackenroth_etal_2010,Har-Shemesh_etal_2012}, were focused on numerical simulations \cite{Li_etal_2018,Harvey_etal_2018} or combined the former approaches \cite{Mackenroth_Holkundkar_2017}. Additional insight could be gained from an improved analytical modeling of the electrons' ponderomotive scattering in a more realistic field shape. 

Here we present an according analytical treatment of electron dynamics in a focused laser field. Furthermore, we apply it exemplary to studies of laser pulse characteristics to be determined from the spatial distribution of an electron bunch scattered off the laser field. Specifically, we study how the laser pulse amplitude can be read off from the maximal scattering angle of the scattered bunch, i.e. the width of the scattered electron distribution. To this end, we demonstrate that due to ponderomotive effects the electron bunch's transverse distribution after scattering will exhibit a cylindrical symmetry with a transverse size determined by the maximal scattering angle. This maximal scattering angle, on the other hand, is directly linked to the ratio of the laser's peak field strength to the electron bunch's peak energy, as was also found in studies of plane wave laser fields \cite{Boca_Florescu_2009,Mackenroth_etal_2010,Mackenroth_DiPiazza_2011,Mackenroth_DiPiazza_2013}. Provided that the latter is well characterized, as is typically feasible for accelerator bunches, the laser intensity can hence be directly read off. Next to this, it can be expected that the scattering angle additionally depends on the laser's spot size. We comment on such a dependence, which is indeed observed in our simulations and point out a path towards this quantity's measurement from spatial distributions of laser-scattered electron bunches.

The paper is organized as follows. After this introductory chapter, we are going to devote chapter 2 to analytically deriving the formula for a laser-scattered electron's final propagation angle. We will find that two distinctly different regimes exist, which we label focus and amplitude dominated, respectively. We will then discuss each of these regimes separately in chapters 3 and 4, respectively. In chapter 5 we will then present numerical benchmarks and simulations of laser-driven electron bunches to confirm our analytical theory. Finally, we will present some considerations for an experimental implementation in chapter 6 and summarize our findings and conclude in chapter 7.
\section{\label{sec:analytics}Derivation of the scattering-angle formula}
\begin{figure}[t]
	\centering
	\includegraphics[width=.6\linewidth]{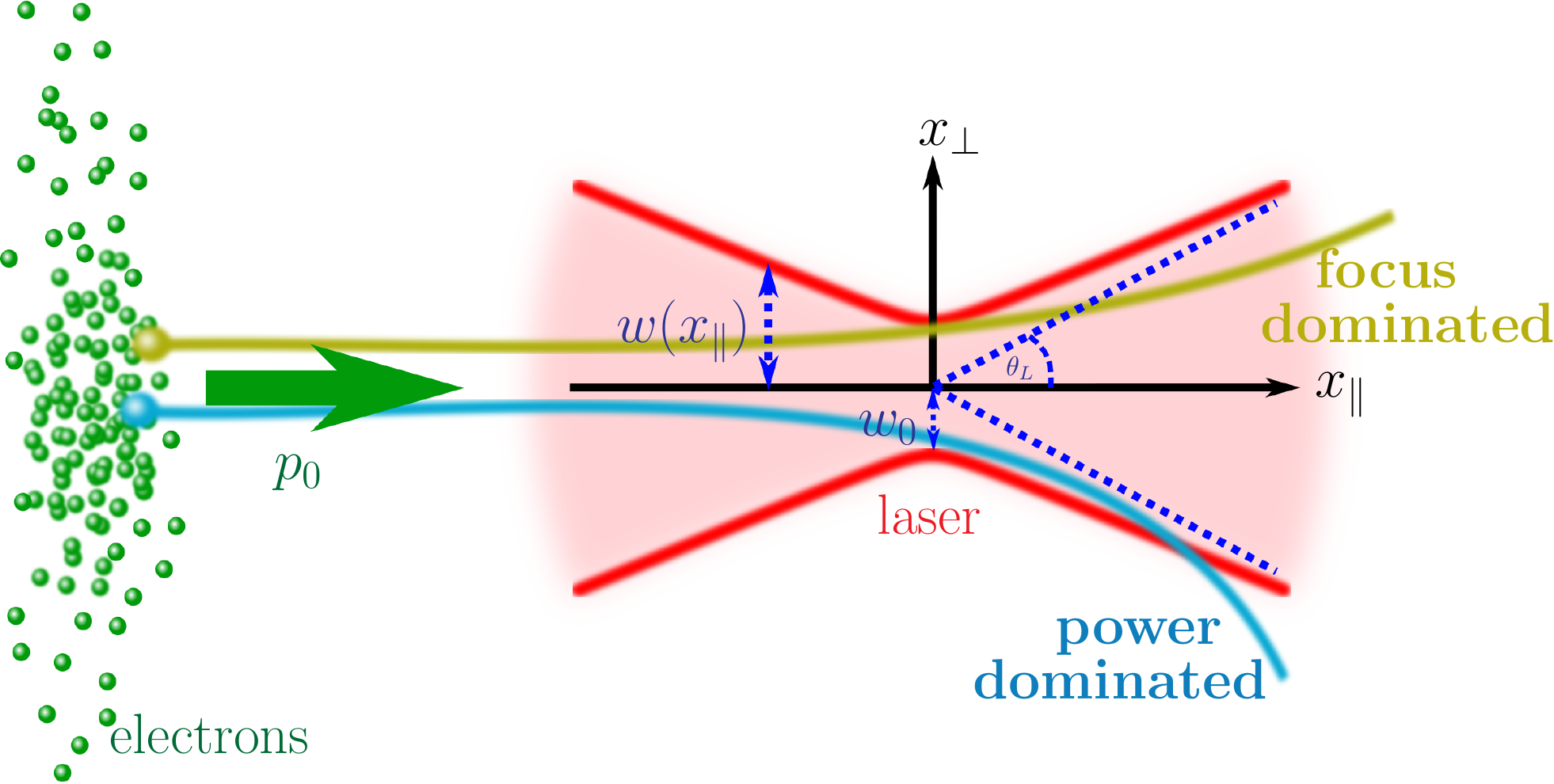}
	\caption{Schematic of the interaction geometry.}
	\label{fig:Schematic}
\end{figure}
We are going to consider the scattering of a bunch of electrons (mass $m$, charge $-e<0$) from a long laser pulse of potential $\bm{A}$ and frequency $\omega$, focused to a Gaussian transverse profile of spot size $w_0$. In accordance with typical laser-particle collision setups, we are going to consider the electron bunch to be collimated and to be dilute, such that space charge effects are negligible and we can consider a single test particle. Next, we choose a reference frame in which the laser pulse propagates along the negative direction $x_\|$ and the electron bunch collides head-on with it. And finally, employing units where the speed of light is $c = 1$, we are going to assume the electron's energy $\varepsilon$ to be much larger than the laser-induced momentum gain $ma_0$, where $a_0=\sqrt{e^2\bm{A}^2/m^2}$ is the dimensionless laser amplitude, also implying $\varepsilon\gg m$, indicating highly relativistic electrons. This results in the observation that the electron's longitudinal momentum $p_\|$ along its initial propagation direction will be largely unaffected by the laser pulse. Furthermore, as argued above, under these assumptions sub-cycle electron oscillations are averaged out in long pulses and the electron dynamics can be modeled by the envelope's ponderomotive effect. Hence, the electron's transverse momentum $\bm{p}_\perp$ to change according to the relativistic generalization of the ponderomotive force equation averaged over the fast oscillations of the scattering laser field \cite{Quesnel_Mora_1998}
\begin{eqnarray}
    \frac{d\bm{p}}{dt} = -\frac{e^2}{2m\overline{\gamma}} \bm{\nabla}\overline{\left|\bm{A}_\perp\right|^2}, \label{eq:relponderomotive}
\end{eqnarray}
where $\bm{A}_\perp$ indicate the two transverse components of the laser's vector potential with respect to $x_\|$, $\overline{\gamma}^2 = 1+\left[\left|\bm{p}_\perp+e\bm{A}_\perp\right|^2 + p_\|^2\right]/m^2$, the overline indicates temporal averaging of the potential's fast oscillations on time scales $\mathcal{O}\left(\omega^{-1}\right)$. As stated above, we consider the electrons' initial energy to be the largest energy scale in the scattering, whence the bunch will be scattered only into small angular deflection by the laser, despite the latter being high-power. We can then approximate the slowly oscillating electron energy to be given by $\overline{\gamma} \approx \sqrt{1+a_0^2+(\varepsilon/m)^2} \approx \textnormal{const}.$, where we additionally respected that the electron's momentum to be only negligibly affected by the laser compared to its initial energy throughout the scattering. In order to estimate the laser's ponderomotive effect on the electron dynamics according to eq. (\ref{eq:relponderomotive}), we need a quantitative model of the laser's perpendicular potential which can be modeled as a Gaussian beam with focal spot radius $w_0$, given by \cite{Quesnel_Mora_1998}
\begin{eqnarray}
    \bm{A}_\perp &= \bm{A}_{\perp,0} \frac{w_0}{w(x_\|)}\exp\left[-\left(\frac{x_\perp}{w(x_\|)}\right)^2 +i\eta\right]\;,\\
    \Rightarrow \overline{\left|\bm{A}_\perp\right|^2} &= A^2_{\perp,0} \left(\frac{w_0}{w(x_\|)}\right)^2\exp\left[-2\left(\frac{x_\perp}{w(x_\|)}\right)^2\right]\;,
\end{eqnarray}
where $w(x_\|)=w_0\sqrt{1+(x_\|/l_R)^2}$ with $l_R = \pi w_0^2/\lambda$ being the Rayleigh length of a laser with wavelength $\lambda = 2\pi/\omega$, $x_\perp = \sqrt{x_{\perp,1}^2+x_{\perp,2}^2}$, with the two coordinate directions $x_{\perp,1}^2,x_{\perp,2}^2$ perpendicular to the laser's propagation direction, $\eta = \omega(t+x_\|)$ is the laser phase and for the sake of simplicity we neglect the Gouy phase and the phase factor accounting for wave front curvature. We see that the potential is independent of the laser's azimuthal angle $\varphi= p_{\perp,2}/p_{\perp,1}$, such that $dp_\varphi/dt \equiv 0$. Then, due to the potential's cylindrical symmetry it is favorable to express the resulting ponderomotive force in cylindrical coordinates with the laser's focal axis as polar direction $x_\|$. The ponderomotive force is then expressed as
\begin{eqnarray}
\frac{dp_\perp}{dt} &= \frac{2e^2A^2_{0}x_\perp}{m\overline{\gamma}}  \left(\frac{w_0}{w^2(x_\|)}\right)^2\exp\left[-2\left(\frac{x_\perp}{w(x_\|)}\right)^2\right] \label{eq:radialmomentum}\\
\frac{dp_\|}{dt} &= \frac{e^2A^2_{0}w_0^2}{m\overline{\gamma}w^3(x_\|)}\left[1+2\frac{x_\perp^2}{w^2(x_\|)}\right]\frac{\partial }{\partial x_\|} w\left(x_\|\right) \nonumber \\
&\times  \exp\left[-2\left(\frac{x_\perp}{w(x_\|)}\right)^2\right].
\end{eqnarray}
For the above specified head-on collision geometry, the electron's initial momentum components will be given by $\bm{p}_{\perp,0} \equiv (0,0)$, $p_{\|,0}\gg m$. Due to $dp_\varphi/dt \equiv 0$ in this configuration the electron momentum's azimuthal angle will remain constant $\varphi = $ const. and the interaction of an initially cylindrically symmetric electron bunch with the laser field will result in a cylindrically symmetric scattering pattern. We can hence confine our analytical analysis to a planar cut through the laser focus containing its focal axis, which we choose to be the $\left(x_{\perp,1},x_\|\right)$-plane. We note that in our calculations the $x_{\perp,1}$-axis does not necessarily coincide with the laser's polarization axis in case of linear polarization, as the ponderomotive scattering is independent of polarization \cite{Quesnel_Mora_1998}. To model a bunch we can then consider single electrons initially distributed according to $\bm{x}_0 = (x_{\|,0},x_{\perp,0},0)$, where the perpendicular displacement from the laser axis $x_{\perp,0}=x_{\perp,1}(t_0)$, with $t_0$ being the initial time, is the impact parameter of conventional scattering theory.

Given that the ultra-relativistic longitudinal momentum will set the largest energy scale in the scattering, we assume the particle's longitudinal velocity to be approximately unaffected by its small radial deflection, whence we can write $x_\|(t) \approx ct$, where for the sake of clarity we made $c$ explicit and as boundary condition we chose that the electron will pass through the focal plane at the time origin $x_\| (t=0) \equiv 0$. This assumption allows us to significantly simplify the analysis of the electron dynamics, as this assumption implicitly also implies $dp_\|/dt \equiv 0$ as well as a prescribed temporal evolution of $w(x_\|)$. Then only the electron's transverse dynamics require a non-trivial solution of Eq.~(\ref{eq:radialmomentum}). As a further crucial simplification, in that equation we approximate the laser's transverse Gaussian profile by a step function, assuming it to be flat within the beam's radial extent $x_\perp \leq w(x_\|)$ and vanishing outside
\begin{eqnarray}\label{eq:transverseapproximation}
    \exp\left[-2\left(\frac{x_\perp}{w(x_\|)}\right)^2\right] = \left\{\begin{array}{cc l}
    1 & \textnormal{if}&  x_\perp \leq w(x_\|)\\
    0 & \textnormal{else}.&
    \end{array}\right.
\end{eqnarray}
We note that by this rough approximation we are going to overestimate the laser's amplitude at least by a factor $\int_0^{w(x_\|)}dx_\perp \exp[-2(x_\perp/w(x_\|))^2]/w(x_\|) \approx 0.75$. We furthermore note that it is crucial that this approximation is made only after the gradient in Eq.~(\ref{eq:relponderomotive}) is evaluated, in order to obtain a non-trivial pre-exponential factor. This pre-exponential, however, dominates the transverse dynamics for $x_\perp \leq w(x_\|)$. We can then write the differential equation for the electron's transverse position as
\begin{eqnarray}
\frac{d^2x_\perp}{dt^2} &= \frac{2e^2A^2_{0} x_\perp}{m^2\overline{\gamma}^2}  \left(\frac{w_0}{w^2(x_\|)}\right)^2 \;, \label{eq:transverseeom}
\end{eqnarray}
i.e., a second order nonlinear differential equation with variable coefficients. Taking into account the presumed time dependence of $x_\| = ct$ this can then be rewritten to
\begin{eqnarray}
\left(1+\left(\frac{ct}{l_R}\right)^2\right)^2\frac{d^2x_\perp}{dt^2} &= \frac{2a_0^2}{\overline{\gamma}^2w_0^2} x_\perp \;. \label{eq:expliciteom}
\end{eqnarray}
We will now be transitioning to the new variables
\begin{eqnarray}
    \tau &= \int_{t_0}^t dt'\ \frac{1}{1+\left(\frac{ct'}{l_R}\right)^2} \;, \label{eq:taudef}\\
    \xi_\perp &= \frac{x_\perp}{\sqrt{1+\left(\frac{ct}{l_R}\right)^2}} \;. \label{eq:xidef}
\end{eqnarray}
In order to evaluate the resulting dynamics, it is instructive to note that $\tau$ has an explicit representation in terms of inverse trigonometric functions
\begin{eqnarray}
    \tau &= l_R\left[\textnormal{arctan}\left(\frac{t}{l_R}\right)+\frac{\pi}2\right] \;. \label{eq:TauLimits}
\end{eqnarray}
In order to decouple the electron and the laser field before scattering completely, we are going to consider $t_0\to-\infty$ such that we find the following asymptotic properties of the rescaled variables (\ref{eq:taudef},\ref{eq:xidef})
\begin{eqnarray}
   \tau &\stackrel{t\to t_0}{\to} 0\ ,\ \tau \stackrel{t\to \infty}{\to} l_R\pi \;, \nonumber\\
\xi &\stackrel{t\to t_0}{\to}  0\ , \ \frac{d}{d\tau}\xi\stackrel{t\to \infty}{\to} - \frac{x_{\perp,0}}{l_R} \;. \nonumber
\end{eqnarray}
Then, with the following differentiation property
\begin{eqnarray}
    \frac{d\xi_\perp}{d\tau} =& \sqrt{1+\left(\frac{ct}{l_R}\right)^2} - \frac{c^2 t \xi_\perp}{l_R^2} \label{eq:secondordereom}
\end{eqnarray}
we see that in the new variables (\ref{eq:taudef},\ref{eq:xidef}) the transverse dynamics are governed by a second order differential equation with constant coefficients
\begin{eqnarray}
    \frac{d^2\xi_\perp}{d\tau^2} &= \left(\frac{2a_0^2}{\overline{\gamma}^2w_0^2} - \frac{1}{l_R^2}\right)\xi_\perp \;, \label{eq:xieom}
\end{eqnarray}
where we put $c=1$ again. For this equation it is reasonable to define the frequency parameter
\begin{eqnarray}
 \Omega = \sqrt{\left|1-2\left(\frac{a_0l_R}{\overline{\gamma}w_0}\right)^2\right|}.
\end{eqnarray}
For the solution of Eq.~(\ref{eq:xieom}) obviously the relative size of the two addends is decisive. Please note here that $a_0$, $w_0$ and $l_R$ are experimentally not independent and ${a_0^2}/{w_0^2}$ can be expressed by the laser pulse' power. Hence for given electron energy $\varepsilon$, the first addend grows only with laser pulse \emph{power}, whereas ${1}/{l_R^2}$ grows just with spatial \emph{focus tightness}.

Equivalently, also the ratio of the two terms $\sqrt2 a_0/\overline{\gamma}$ and $w_0/l_R$ is decisive. Respecting that for $\varepsilon \gg m a_0$, the ratio $\sqrt2 a_0/\overline{\gamma}$ determines an electron's scattering angle $\theta$ from the laser field, whereas $w_0/l_R$ gives the divergence angle $\theta_{\rm{L}}$ of a Gaussian laser beam in the farfield ($ |x_\| | \gg l_R$, see blue dotted line in Fig.~\ref{fig:Schematic}). Hence, one may distinguish two cases, cf.~Fig.~\ref{fig:Schematic}: (i) For $\sqrt2 a_0/\overline{\gamma} = \leq w_0/l_R $ the electron will typically be scattered by the laser pulse to angles within the laser's focussing cone, $\theta \leq \theta_{\rm{L}}$ (s. also below). Furthermore, Eq.~(\ref{eq:xieom}) indicates that in this case $(d^2\xi_\perp/d\tau^2)/\xi_\perp \leq 0$. Since the relation between rescaled transverse position and acceleration in this case is dominated by the focusing of the laser pulse, we term this regime the \textit{focus dominated}. (ii) On the other hand, for $\sqrt2 a_0/\overline{\gamma} \geq w_0/l_R$, the electron will typically be scattered to angles \emph{outside} of the laser's focal cone, $\theta > \theta_{\rm{L}}$, and it holds $(d^2\xi_\perp/d\tau^2)/\xi_\perp \geq 0$. Based on the same argument as before, the laser pulse power dominates and we term this regime \textit{amplitude dominated}.

Before we turn to quantitative solutions of Eq.~(\ref{eq:xieom}) in these two cases, we note that interestingly, under our rough approximations it appears that there exists an \emph{equilibrium} between the laser's focusing and intensity on the one hand and the electron's initial energy on the other: For a balanced relation $w_0/l_R = \sqrt{2} a_0/\overline{\gamma}$ an electron initially propagating parallel to the laser axis will be driven from this propagation state by the strongly simplified second order differential equation $d^2\xi_\perp/d\tau^2\equiv 0$. Hence, the electron will experience linear deflection by the laser and, in this case, we read off from Eq.~(\ref{eq:xieom}) the general solution for the equilibrium transverse displacement
\begin{eqnarray}
    \fl x^\textnormal{eq}_\perp(t) &= \sqrt{1+\left(\frac{t}{l_R}\right)^2}\left(\alpha^\textnormal{eq} + \beta^\textnormal{eq} l_R\left[\textnormal{arctan}\left(\frac{t}{l_R}\right)+\frac{\pi}2\right]\right)\\
    \fl \frac{d}{dt} x^\textnormal{eq}_\perp(t) &= v_\perp^\textnormal{eq}(t)= \frac{t\left(\alpha^\textnormal{eq} + \beta^\textnormal{eq} l_R\left[\textnormal{arctan}\left(\frac{t}{l_R}\right)+\frac{\pi}2\right]\right)}{l_R^2\sqrt{1+\left(\frac{t}{l_R}\right)^2}} + \frac{\beta^\textnormal{eq}}{\sqrt{1+\left(\frac{t}{l_R}\right)^2}}.
\end{eqnarray}
From the boundary conditions $t_0=-\infty$, $x^\textnormal{eq}_\perp(t) = x_{\perp,0}$, $v^\textnormal{eq}_\perp(t_0)=0$ we find the coefficients to be given by $\alpha^\textnormal{eq} \equiv 0$, $\beta^\textnormal{eq}\equiv x_{\perp,0}/l_R$, such that in the equilibrium case the transverse electron dynamics are given by
\begin{eqnarray}
    x^\textnormal{eq}_\perp(t) &= x_{\perp,0} \sqrt{1+\left(\frac{t}{l_R}\right)^2}\left[\textnormal{arctan}\left(\frac{t}{l_R}\right)+\frac{\pi}2\right] \label{eq:equilibriumsolution}
\end{eqnarray}
Since we are interested in the electron's scattering, we need to connect these dynamics to the instantaneous propagation angle $\theta (t)$. Since we assumed $v_\| = 1$ in the above derivation, this angle is numerically equivalent to the electron's transverse velocity 
\begin{eqnarray}
\theta (t) = v_\perp(t) = \frac{x_{\perp,0}}{l_R\sqrt{1+\left(\frac{t}{l_R}\right)^2}}\left[\frac{t}{l_R}\left(\textnormal{arctan}\left(\frac{t}{l_R}\right)+\frac{\pi}2\right)+1\right], \label{eq:EQ_transversevelocity}
\end{eqnarray}
according to the above dynamics.

Next, we need to distinguish the following two cases: (1) For an initial transverse displacement $x_{\perp,0}\leq w_0/\pi$ it will always be $x^\textnormal{eq}_\perp(t)\leq w(x_\|)$, i.e., the electron will remain inside the laser's focal volume. In this case the electron's maximal scattering angle will be reached at asymptotic times $t\to \infty$ and be given by $\theta(t\to\infty) = \pi x_{\perp,0}/l_R\leq w_0/l_R$. (2) For $x_{\perp,0}\geq w_0/\pi$ the electron will leave the laser's focal volume at a maximum propagation time 
\begin{eqnarray}
t_\textnormal{max} &=l_R\tan\left(\frac{w_0}{x_{\perp,0}}-\frac{\pi}2\right), \label{eq:EQ_maximaltime}
\end{eqnarray}
propagating towards an angle
\begin{eqnarray}
\theta (t_\textnormal{max}) = \frac{w_0}{l_R} \left[\frac{x_{\perp,0}}{w_0}\sin\left(\frac{w_0}{x_{\perp,0}}\right)-\cos\left(\frac{w_0}{x_{\perp,0}}\right)\right]. \label{eq:EQ_maximalangle}
\end{eqnarray}
From the derivative $d/dx_{\perp,0}\theta (t_\textnormal{max})$ we find that the maximal scattering angle is reached at an initial transverse displacement satisfying the equation 
\begin{eqnarray}
\tan(v)=\frac{v}{(1-v^2)},
\end{eqnarray}
where $v:=w_0/x_{\perp,0}$. While this equation is not analytically solvable, it is always maximized at $x_{\perp,0}^\textnormal{peak}\approx w_0/2.74$, irrespective of the laser spot size $w_0$. Inserting this value, together with its defining equation, into (\ref{eq:EQ_maximalangle}), one finds that the maximal scattering angle in the equilibrium case is 
\begin{eqnarray}
\theta^\textnormal{max} = \theta \left(t_\textnormal{max}\left[x_{\perp,0}^\textnormal{peak}\right]\right) \approx 1.06\frac{w_0}{l_R}\;,
\end{eqnarray}
being slightly out of the laser's focussing cone.

Naturally, it has to be assumed that the laser's nontrivial transverse field distribution within the beam will not permit this symmetry to be perfect but for not too tight focusing it might still be approximately observable.
\section{\label{sec:focusdominated}Focus dominated ponderomotive scattering}
In the case $w_0/l_R = \lambda/\pi w_0 \geq \sqrt{2}a_0/\overline{\gamma}$ what is equivalent to a too low laser power compared to $\overline{\gamma}$, Eq.~(\ref{eq:xieom}) turns into a harmonic oscillator equation which is canonically solved by
\begin{eqnarray}
    \xi_\perp(t) &= \alpha \sin\left[\frac{\Omega}{l_R} \tau\right] + \beta \cos\left[\frac{\Omega}{l_R} \tau\right] \label{eq:FD_solution}. 
\end{eqnarray}
Consequently, the electron's perpendicular coordinate is given by
\begin{eqnarray}
    x_\perp (t) &= \alpha \sqrt{1+\left(\frac{t}{l_R}\right)^2}\sin\left[\Omega \left(\textnormal{arctan}\left(\frac{t}{l_R}\right)+\frac{\pi}2\right)\right] \nonumber \\
    &+ \beta \sqrt{1+\left(\frac{t}{l_R}\right)^2}\cos\left[\Omega\left(\textnormal{arctan}\left(\frac{t}{l_R}\right)+\frac{\pi}2\right)\right].
\end{eqnarray}
The constant coefficients in the focus dominated regime of Eq.~(\ref{eq:FD_solution}) are again obtained by imposing boundary conditions. Analogously to the above discussion, it is favorable to first impose the assumption of vanishing initial transverse velocity, i.e., that the electron initially propagates along the laser's propagation axis
\begin{eqnarray}
    \fl \left.v_\perp\right|_{t\to t_0} &= \frac{\alpha}{l_R}\sin\left[\Omega \left(\textnormal{arctan}\left(\frac{t}{l_R}\right)+\frac{\pi}2\right)\right] + \frac{\beta}{l_R}\cos\left[\Omega \left(\textnormal{arctan}\left(\frac{t}{l_R}\right)+\frac{\pi}2\right)\right].
\end{eqnarray}
We hence see that $\beta\equiv0$ is required to fulfill $v_\perp(t_0)=0$. In order to fix the prefactor $\alpha$ we need to impose the electrons initial transverse displacement $x_\perp(t_0) = x_{\perp,0}$. Since we chose $t_0=-\infty$ formally $x_\perp(t_0)$ is undefined. We can, however, circumvent this complication by considering the limit $t\to t_0$ and taking into account $\textnormal{arctan}(x\to-\infty)\to -\pi/2 - 1/x$ as well as $\sin(x\to0)\to x$. Then, we find for the electron's perpendicular coordinate at times approaching $t_0$
\begin{eqnarray}
    x_\perp (t\to t_0) &\to \alpha \Omega \equiv x_{\perp,0}.
\end{eqnarray}
Consequently, the free parameter is $\alpha = x_{\perp,0}/\Omega$ and the transverse electron dynamics are explicitly given by
\begin{eqnarray}
    x_\perp (t) &= \frac{x_{\perp,0}}{\Omega} \sqrt{1+\left(\frac{t}{l_R}\right)^2}\sin\left[\Omega \left(\textnormal{arctan}\left(\frac{t}{l_R}\right)+\frac{\pi}2\right)\right]. \label{eq:FD_explicitsolution}
\end{eqnarray}
It is interesting to note that Eq.~(\ref{eq:equilibriumsolution}) obviously is the $\Omega\to0$ expansion of this general solution. This could have been expected as the equilibrium condition $w_0/l_R = \sqrt{2} a_0/\overline{\gamma}$ precisely translates to $\Omega\equiv0$, which is superficially divergent in (\ref{eq:FD_explicitsolution}). The electron's transverse velocity is then given by
\begin{eqnarray}
    \fl v_\perp (t) =& \frac{x_{\perp,0}}{\Omega \sqrt{l_R^2+t^2}} \left(\frac{t}{l_R}\sin\left[\Omega \left(\textnormal{arctan}\left(\frac{t}{l_R}\right)+\frac{\pi}2\right)\right] \right.\nonumber \\
    &\left.+ \Omega\cos\left[\Omega \left(\textnormal{arctan}\left(\frac{t}{l_R}\right)+\frac{\pi}2\right)\right] \right). \label{eq:FD_transversevelocity}
\end{eqnarray}
%
%
%
Analogous to the above discussion, we obtain the electron's propagation angle as a function of time by studying $v_\perp(t)$. First, it is instructive to compare this angle to the laser's divergence
\begin{eqnarray}
    \tan \theta_L := \left.\frac{w(x_\|)}{x_\|}\right|_{x_\| \to \infty} = \frac{w_0}{l_R} = \frac{\lambda}{\pi w_0} \;.
\end{eqnarray}
Any electron scattered into angles $\theta (t)\geq \theta_L$ will leave the focal volume at some time. To estimate this time, we note that the ratio of the electron's perpendicular position to the beam's radius to be given by
\begin{eqnarray}
    \frac{x_\perp(t)}{w(x_\|)} &= \frac{x_{\perp,0}}{w_0 \Omega} \sin\left[\Omega \left(\textnormal{arctan}\left(\frac{t}{l_R}\right)+\frac{\pi}2\right)\right] \;, \label{eq:FD_transverseposition}
\end{eqnarray}
whence we conclude that, in accordance with the above discussion of the focus dominated regime, the electron will never exit the laser beam for initial displacements smaller than the maximal value
\begin{eqnarray}\label{eq:FD_maximaldisplacement}
 x_{\perp,0}\leq x_{\perp,0}^\textnormal{max} := w_0 \Omega \;.
\end{eqnarray}
(recall that it always holds $\Omega\leq1$ in the focus dominated regime). Consequently, in the focus dominated regime and for electrons staying inside the laser's focal volume for all time, i.e., for $x_{\perp,0} \in \left[0,w_0 \Omega\right]$, the final scattering angle $\theta_f$ can be determined to be
\begin{eqnarray}
\tan \theta_f &:= \tan \theta(t\to\infty)= \frac{x_{\perp,0}}{\Omega l_R} \sin\left[\pi \Omega\right] \leq \frac{w_0}{l_R} \;. \label{eq:FD_scatteringangle}
\end{eqnarray}
In this equation, the fact that $\theta_f(x_{\perp,0}=0)=0$ is consistent with the well-known fact that when propagating exactly on axis a particle will not experience any ponderomotive scattering due to the potential envelope's radial symmetry around this axis. On the other hand, the fact that the scattering angles are linear in the initial displacement $x_{\perp,0}$ is an unphysical artifact of modeling the laser's transverse intensity profile as flat within its beam diameter in Eq.~(\ref{eq:transverseapproximation}). Including the exponential factor in that equation will most likely introduce a non-trivial $x_{\perp,0}$ dependence in Eq.~(\ref{eq:FD_scatteringangle}) but also complicates Eq.~(\ref{eq:transverseeom}) to an exponentially nonlinear second order differential equation. 

\begin{figure}[t]
	\centering
	\includegraphics[width=.6\linewidth]{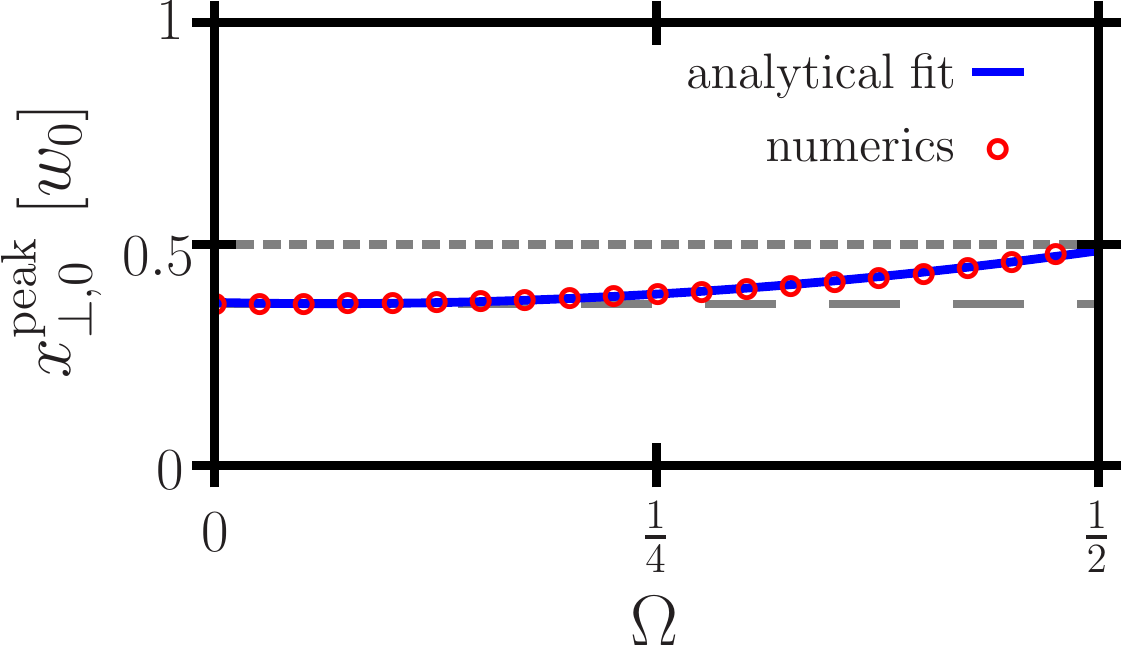}
	\caption{Transverse displacement $x_{\perp,0}^\textnormal{peak}$ resulting in maximal electron scattering in the focus dominated regime in units of the laser spot size $w_0$ as function of $\Omega \in [0,1/2]$. The two limits $x_{\perp,0}^\textnormal{peak}(\Omega=0)\approx w_0/2.74$ (lower gray line, dashed) and $x_{\perp,0}^\textnormal{peak}(\Omega=1/2)=w_0/2$ (upper gray line, dotted) are well reproduced.}
	\label{fig:Figure6}
\end{figure}

For $x_{\perp,0}\geq w_0 \Omega$, on the other hand, we need to distinguish the cases $\Omega \leq 1/2$ and $\Omega > 1/2$, respectively. In the former case, the electron will reach its closest approach to the focal boundary $w(x_\|)$ only for $t\to\infty$. In this case the sine in Eq.~(\ref{eq:FD_transverseposition}) will not reach its maximum and $x_\perp(t)/w(x_\|)$ hence be a monotonically increasing function of time. Consequently, the electron will only reach its nearest approach to the focal volume's boundary at asymptotic times $t\to\infty$. Hence, in this regime the electron will stay inside the laser focus only if it started at initial transverse displacements smaller than 
\begin{eqnarray}
 x_{\perp,0} \leq x_{\perp,0}^\textnormal{max}= \frac{w_0 \Omega}{\sin[\pi \Omega]} \;.
\end{eqnarray}
We can then define a unique time at which the electron leaves the laser's focal region by solving eq. (\ref{eq:FD_transverseposition}) for $x_\perp(t)/w(x_\|)\equiv 1$, resulting in
\begin{eqnarray}
t_\textnormal{max} &=l_R\tan\left(\frac{\arcsin\left[\frac{w_0 \Omega}{x_{\perp,0}}\right]}{\Omega}-\frac{\pi}2\right) \;. \label{eq:FD_maximaltime}
\end{eqnarray}
Consequently, neglecting re-entry into the focal volume, for an electron leaving the laser's focal volume the final scattering angle will not be determined by the limit $t\to\infty$, as the electron will not be accelerated up to that time, but only until $t_\textnormal{max}$, according to eq. (\ref{eq:FD_maximaltime}). Hence, the electrons' scattering angle will be given by $\tan \theta \left(t_\textnormal{max}\right) =\left( t_\textnormal{max} w_0/l_R+ \sqrt{x_{\perp,0}^2-\left(w_0 \Omega\right)^2} \right)/\sqrt{l_R^2+t_\textnormal{max}^2} $. Inserting (\ref{eq:FD_maximaltime}) into this expression, we find the maximal scattering angle for electrons leaving the focal volume to be given by
\begin{eqnarray}
\fl \tan \theta \left(t_\textnormal{max}\right) &= \frac{w_0}{l_R} \left(\sqrt{\left(\frac{x_{\perp,0}}{w_0 }\right)^2-\Omega ^2}\sin\left(g(x_{\perp,0})\right) - \cos\left(g(x_{\perp,0})\right) \right) \;, \label{eq:FD_maxscatteringangle}
\end{eqnarray}
where we defined the argument of the trigonometric functions $g(x_{\perp,0}):=\arcsin\left[w_0 \Omega/x_{\perp,0}\right]/\Omega$, satisfying $g'(x_{\perp,0}):=-1/x_{\perp,0}\sqrt{(x_{\perp,0}/w_0)^2 - \Omega^2}$, i.e., monotonically falling in the regime $\Omega \leq 1/2$ where it holds $x_{\perp,0}\in[w_0\Omega/\sin[\pi \Omega],\infty]$ and hence $g(x_{\perp,0})\in[\pi,0]$. We again recognize (\ref{eq:EQ_maximalangle}) to be the $\Omega\to0$ expansion of (\ref{eq:FD_maxscatteringangle}). Comparing (\ref{eq:FD_scatteringangle}) and (\ref{eq:FD_maxscatteringangle}) we see that for $\Omega\leq1/2$ at for an initial displacement $x_{\perp,0} = x_{\perp,0}^\textnormal{max}$, right at the boundary between the regimes where an electron stays inside or leaves the laser's focal volume, it holds $t_\textnormal{max}(x_{\perp,0}\to x_{\perp,0}^\textnormal{max})) \to \infty$ such that $\tan \theta(t_\textnormal{max}(x_{\perp,0}^\textnormal{max})) = w_0/l_R =  x_{\perp,0}^\textnormal{max}\sin\left[\pi \Omega\right]/\Omega l_R$. Hence, the analytical result predicts a continuous distribution of scattering angles, irrespective of whether the electron leaves or remains inside the focal volume. Furthermore, we can study, whether for a particular initial displacement, there exists a maximum in the scattering angle. To this end, we note that for electrons leaving the focal volume, the scattering angle changes as a function of initial transverse displacement according to 
\begin{eqnarray}
\fl &\frac{d}{dx_{\perp,0}} \tan \theta \left(t_\textnormal{max}\right) = -\frac{w_0\sin\left(g(x_{\perp,0})\right)}{x_{\perp,0}l_R} \left[\frac{\frac{x_{\perp,0}^2} {w_0^2}-1}{\sqrt{\left(\frac{x_{\perp,0}}{w_0}\right)^2 - \Omega^2}} -\cot\left(g(x_{\perp,0})\right) \right] \;. \label{eq:FD_maxscatteringanglederivative}
\end{eqnarray}
From the above considerations on the domain of $g(x_{\perp,0})$ we conclude that the prefactor is strictly positive in the regime $\Omega \leq 1/2$, whence the scattering angle can only exhibit a maximum if the term in brackets vanishes. In analogy to the discussion following Eq.~(\ref{eq:EQ_maximalangle}) we find that for finite $\Omega \in [0,1/2]$ the maximal scattering angle will be determined by the more complicated equation
\begin{eqnarray}
\tan\left(\frac{\arcsin(\Omega v(\Omega))}{\Omega}\right) = \frac{v\sqrt{1-(\Omega v(\Omega))^2}}{1-v^2(\Omega)} \;,
\end{eqnarray}
where we recall $v(\Omega)=w_0/x_{\perp,0}(\Omega)$. This equation is again not solvable analytically, but we can expect that for small $\Omega \leq 1/2$ the solution $v(\Omega)$ will scale quadratically in $\Omega$ at most. Solving the defining equation for $v(\Omega)$ we find again a behavior independent of $w_0$ and the two limits $x_{\perp,0}^\textnormal{peak}(\Omega=0)\approx w_0/2.74$ (s. discussion before (\ref{eq:EQ_maximalangle}) and following) and $x_{\perp,0}^\textnormal{peak}(\Omega=1/2)=w_0/2$ (s. below) well reproduced (s. fig. \ref{fig:Figure6}). Analytically, the peak position is well approximated by $x_{\perp,0}^\textnormal{peak}(\Omega)\approx w_0(0.37 - 8\times 10^-2 \Omega + 0.64 \Omega^2)$, i.e., a modified quadratic scaling as conjectured above.
%

Additionally, it is important to note that for large initial transverse displacements $x_{\perp,0}\to\infty$, the exit time diverges as $t_\textnormal{max}\to-\infty$. Inserting this relation into Eq.~(\ref{eq:FD_maxscatteringangle}), on the other hand, we see, that for large initial transverse displacements the scattering angle vanishes $\tan \theta_f (x_{\perp,0}\to \infty) = 0$. Hence, it appears that for $\Omega \leq 1/2$ the maximum scattering angle will be reached for initial displacements $x_{\perp,0} \approx x_{\perp,0}^\textnormal{max}$. 

For the latter case $\Omega > 1/2$, on the other hand, the electron will reach its maximal approach to the focal boundary at a finite time, and then retreat from it again. As a consequence, the maximal initial transverse displacement at which an electron will still remain inside the laser's focal volume throughout the scattering is given by (\ref{eq:FD_maximaldisplacement}) and (\ref{eq:FD_maximaltime}) will possibly have multiple solutions. Of these we naturally have to consider the earlier time, as we do not wish to consider re-entry into the focal volume. Comparing (\ref{eq:FD_scatteringangle}) and (\ref{eq:FD_maxscatteringangle}) we find that in this case of $\Omega > 1/2$ the analytical result does not predict a continuous distribution of scattering angles for varying initial displacements $x_{\perp,0}$ but is discontinuous at $x_{\perp,0}^\textnormal{max}$. To see this, it is sufficient to note that in this regime $x_{\perp,0}^\textnormal{max}\sin\left[\pi \Omega\right]/\Omega l_R = \sin\left[\pi \Omega\right] w_0/l_R$. On the other hand, the time at which the electron leaves the laser's focal region is finite even for $x_{\perp,0} = x_{\perp,0}^\textnormal{max}$ and given by $t_\textnormal{max}(x_{\perp,0}^\textnormal{max})) = l_R \tan((1/\Omega - 1)\pi/2 )$. We hence conclude that in the regime $\Omega>1/2$ the electron will leave the laser's focal volume always at times $t_\textnormal{max}\leq t_\textnormal{max}(x_{\perp,0}^\textnormal{max})$. The corresponding scattering angle is given by $\tan \theta \left(t_\textnormal{max}(x_{\perp,0}^\textnormal{max})\right) = \sin\left[\pi/2 (1/\Omega-1)\right] w_0/l_R \leq x_{\perp,0}^\textnormal{max}\sin\left[\pi \Omega\right]/\Omega l_R$. Consequently, in this regime the maximal scattering angle will be reached for initial transverse displacements of $x_{\perp,0}^\textnormal{peak}=x_{\perp,0}^\textnormal{max}$. The derived discontinuous behavior is an artifact of our model assumption to neglect reentry of the electrons into the focal volume. We will see below, however, that the analytical formulas still capture the main qualitative features of the ponderomotive scattering. Furthermore, from evaluating (\ref{eq:FD_maxscatteringanglederivative}) we find that the scattering angle is monotonically decreasing for increasing $x_{\perp,0}$, in contrast to the above case $\Omega \leq 1/2$, whence it is obvious that the maximal scattering angle to be reached for electrons with initial displacement $x_{\perp,0}=x_{\perp,0}^\textnormal{max}$. Furthermore, it also holds $\tan \theta_f (x_{\perp,0}\to\infty)\to0$ even in the regime $\Omega > 1/2$.

Next, it is instructive to study the scaling properties of the scattering angles. Since we assumed the electron's initial energy to be the dominating energy scale in the problem, we can always assume $a_0/\overline{\gamma}\ll 1$ and expand all solutions in this small parameter. We then find $\Omega \approx 1 - (a_0l_R / \overline{\gamma}w_0)^2$ and hence
\begin{eqnarray}
    \tan \theta_f \approx \pi \frac{x_{\perp,0}}{\lambda} \left(\frac{a_0}{\overline{\gamma}}\right)^2,
\end{eqnarray}
for scattering being determined by the focusing. We hence see that the final scattering angle should scale as $\tan \theta_f \sim (a_0/\overline{\gamma})^2$. We thus see that for lower energies $\varepsilon$ and larger amplitudes $a_0$ the scattering becomes more pronounced, as it has to be. Consequently, the opening angle of the scattered electron bunch can obviously serve to infer information about the scattering laser's intensity. We note that the scaling of the scattering angle with changing $w_0$ cannot be derived from the above discussion, due to the strongly simplifying assumption of a step-like transverse laser profile, which masks the nature of $w_0$ as the exponential decay length of the transverse profile.

\section{\label{sec:amplitudedominated}Amplitude dominated ponderomotive scattering}
In the amplitude dominated regime $w_0/l_R \leq \sqrt{2}a_0/\overline{\gamma}$, in contrast to the discussion of the previous section, Eq.~(\ref{eq:xieom}) turns into an exponentially accelerated second order differential equation which is canonically solved by
\begin{eqnarray}
\xi_\perp(t) &= \alpha \sinh\left[\frac{\Omega}{l_R} \tau\right] + \beta \cosh\left[\frac{\Omega}{l_R} \tau\right] \label{eq:AD_solution} \;. 
\end{eqnarray}
In analogy to the analysis in the focus dominated regime, from imposing boundary conditions on this general solution, the transverse dynamics turn out to be given by
\begin{eqnarray}
x_\perp (t) &= \frac{x_{\perp,0}}{\Omega} \sqrt{1+\left(\frac{t}{l_R}\right)^2}\sinh\left[\Omega \left(\textnormal{arctan}\left(\frac{t}{l_R}\right)+\frac{\pi}2\right)\right] \label{eq:AD_explicitsolution} \;,
\end{eqnarray}
where we recognize the only difference to Eq.(\ref{eq:FD_explicitsolution}) to be the hyperbolic sine replacing its trigonometric counterpart. This replacement, however, implies the possibility for unbounded growth of the perpendicular coordinate, whence many physical properties of the solution will change. The electron's transverse velocity in this case is given by
\begin{eqnarray}
\fl v_\perp (t) &= \frac{x_{\perp,0}}{\Omega \sqrt{l_R^2+t^2}} \left(\frac{t}{l_R}\sinh\left[\Omega \left(\textnormal{arctan}\left(\frac{t}{l_R}\right)+\frac{\pi}2\right)\right] \right.\nonumber \\
\fl &\left.+ \Omega\cosh\left[\Omega \left(\textnormal{arctan}\left(\frac{t}{l_R}\right)+\frac{\pi}2\right)\right] \right) \;. \label{eq:AD_transversevelocity}
\end{eqnarray}
%
Next, we again wish to establish a connection between the electron's transverse velocity and its scattering angle. We recall again that within the limits of the above derivation the transverse velocity is numerically equivalent to the electron's instantaneous propagation angle $v_\perp(t) = v_\perp(t) /v_\|(t) = \theta (t)$. In order to estimate the time at which an electron will leave the laser's focal volume, we again consider the electron's perpendicular position in units of the laser beam's radius
\begin{eqnarray}
\frac{x_\perp(t)}{w(x_\|)} &= \frac{x_{\perp,0}}{w_0 \Omega} \sinh\left[\Omega \left(\textnormal{arctan}\left(\frac{t}{l_R}\right)+\frac{\pi}2\right)\right] \;. \label{eq:AD_transverseposition}
\end{eqnarray}
Here we note that in the amplitude dominated regime $\Omega$ is not bounded but can, in principle, grow arbitrarily large. Hence, the electron will only remain within the laser's focal volume provided
\begin{eqnarray}
\label{eq:AD_MaximumInitialDisplacement}
x_{\perp,0}\leq x_{\perp,0}^\textnormal{max} := \frac{w_0 \Omega}{\sinh[\pi \Omega]}.
\end{eqnarray}
Provided this prerequisite is satisfied, the final scattering angle is given by
\begin{eqnarray}
\tan \theta_f &:= \tan \theta(t\to\infty)= \frac{x_{\perp,0}}{\Omega l_R} \sinh\left[\pi \Omega \right]. \label{eq:AD_scatteringangle}
\end{eqnarray}
Combining Eqs.~(\ref{eq:AD_MaximumInitialDisplacement},\ref{eq:AD_scatteringangle}) we find that the maximal scattering angles of electrons remaining within the laser's focal volume is reached by electrons with initial transverse displacement $x_{\perp,0}^\textnormal{max}$ and is given by $\tan \theta_f^\textnormal{max} = w_0/l_R$, i.e., the laser's divergence angle. This result, however, had again to be expected, as $\theta_L = w_0/l_R$ is the maximal deflection angle for which an electron can asymptotically remain inside the laser's focal volume.

For $x_{\perp,0}\geq x_{\perp,0}^\textnormal{max}$, on the other hand, independent of the value of $\Omega$, the electron will leave the laser's focal volume. In close analogy to the discussion connected to Eqs.~(\ref{eq:FD_maximaltime},\ref{eq:FD_maxscatteringangle}) the maximal propagation time is given by 
\begin{eqnarray}
t_\textnormal{max} &=l_R\tan\left(\frac{\textnormal{arcsinh}\left[\frac{w_0 \Omega}{x_{\perp,0}}\right]}{\Omega}-\frac{\pi}2\right). \label{eq:AD_maximaltime}
\end{eqnarray}
An electron leaving the laser's focal volume at this time will propagate towards an angle $\tan \theta \left(t_\textnormal{max}\right) = \left( t_\textnormal{max} w_0/l_R + \sqrt{x_{\perp,0}^2+\left(w_0 \Omega\right)^2} \right)/\sqrt{l_R^2+t_\textnormal{max}^2}$. Upon insertion of (\ref{eq:AD_maximaltime}) into this expression we find the maximal scattering angle for electrons leaving the focal volume in the amplitude dominated regime to be formally equivalent to (\ref{eq:FD_maxscatteringangle})
\begin{eqnarray}
\fl \tan \theta \left(t_\textnormal{max}\right) &= \frac{w_0}{l_R}\left(\sqrt{\left(\frac{x_{\perp,0}}{w_0}\right)^2+\Omega^2}\sin\left(\tilde{g}(x_{\perp,0})\right) -\cos\left(\tilde{g}(x_{\perp,0})\right)\right), \label{eq:AD_maxscatteringangle}
\end{eqnarray}
albeit with an altered argument of the trigonometric functions in the amplitude dominated regime $\tilde{g}(x_{\perp,0}):=\textnormal{arcsinh}\left[w_0 \Omega / x_{\perp,0}\right]/\Omega$, satisfying $g'(x_{\perp,0}):=-1/x_{\perp,0}\sqrt{(x_{\perp,0}/w_0)^2 + \Omega^2}$. In analogy to the focus dominated regime from $x_{\perp,0}\in[w_0\Omega/\sinh[\pi \Omega],\infty]$ it follows $g(x_{\perp,0})\in[\pi,0]$. 

Analogous to the case $\Omega\leq 1/2$ in the focus dominated regime, from comparing (\ref{eq:AD_scatteringangle}) and (\ref{eq:AD_maxscatteringangle}) we find that in the amplitude dominated regime for an initial displacement $x_{\perp,0} = x_{\perp,0}^\textnormal{max}$, right at the boundary between the regimes where an electron stays inside or leaves the laser's focal volume, it holds $\tan \theta(t_\textnormal{max}(x_{\perp,0}^\textnormal{max})) = w_0/l_R =  x_{\perp,0}^\textnormal{max}\sinh\left[\pi \Omega\right]/\Omega l_R$. Hence, the analytical result again predicts a continuous distribution of scattering angles, irrespective of whether the electron leaves or remains inside the focal volume. 

\begin{figure}[t]
	\centering
	\includegraphics[width=.6\linewidth]{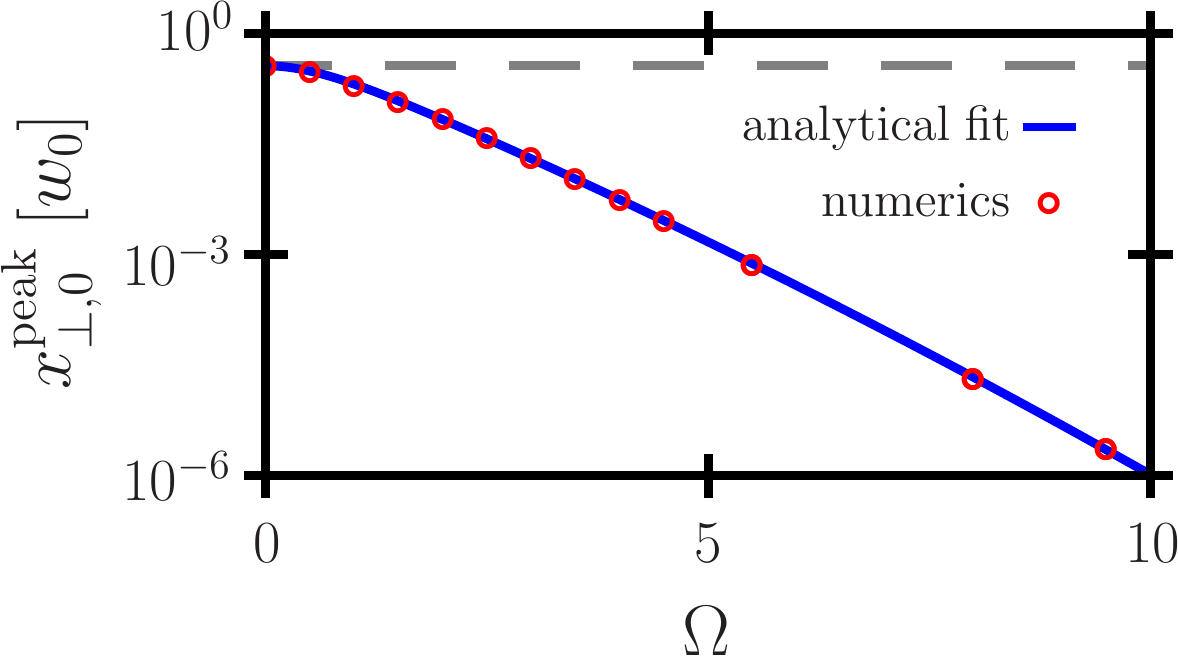}
	\caption{Transverse displacement $x_{\perp,0}^\textnormal{peak}$ resulting in maximal electron scattering in the amplitude dominated regime in units of the laser spot size $w_0$ as function of $\Omega \in [0,10]$. The limit $x_{\perp,0}^\textnormal{peak}(\Omega=0)\approx w_0/2.74$ (dashed gray line) is well reproduced.}
	\label{fig:Figure7}
\end{figure}
Repeating the above search for the maximum scattering angle as a function of initial transverse displacement, we find the derivative of (\ref{eq:AD_maxscatteringangle})
\begin{eqnarray}
\fl &\frac{d}{dx_{\perp,0}} \tan \theta \left(t_\textnormal{max}\right) = -\frac{w_0\sin\left(\tilde{g}(x_{\perp,0})\right)}{x_{\perp,0}l_R} \left[\frac{\frac{x_{\perp,0}^2} {w_0^2}-1}{\sqrt{\left(\frac{x_{\perp,0}}{w_0}\right)^2 + \Omega^2}} -\cot\left(\tilde{g}(x_{\perp,0})\right) \right]. \label{eq:AD_maxscatteringanglederivative}
\end{eqnarray}
From the above considerations on the domain of $\tilde{g}(x_{\perp,0})$ we conclude that in the amplitude dominated regime the maximal scattering angle will be determined by the equation
\begin{eqnarray}
\tan\left(\frac{\textnormal{arcsinh}( \Omega v(\Omega))}{\Omega}\right)=\frac{v\sqrt{1-(\Omega v(\Omega))^2}}{1-v^2(\Omega)},
\end{eqnarray}
where we recall $v(\Omega)=w_0/x_{\perp,0}(\Omega)$. This equation is again not solvable analytically, but since $\Omega$ can grow arbitrarily large in the amplitude dominated regime, we can expect that the solution $v(\Omega)$ will scale inversely proportional to a hyperbolic trigonometric function converging to a finite value for $\Omega \to 0$, a property is particular to the hyperbolic cosine. Solving then the defining equation for $v(\Omega)$ we find again a behavior independent of $w_0$ and the limit $x_{\perp,0}^\textnormal{peak}(\Omega=0)\approx w_0/2.74$ well reproduced (s. fig. \ref{fig:Figure7}). Analytically, the peak position is well approximated by $x_{\perp,0}^\textnormal{peak}(\Omega)\approx w_0/v(\Omega=0)/\cosh(3.8\times 10^{-2} + 1.12\, \Omega + 2.2 \times 10^{-2} \Omega^2)$, i.e., an inverse hyperbolic cosine with a quadratic argument scaling as conjectured above.

Finally, in contrast to the focus dominated regime, in the amplitude dominated regime we cannot expand the parameter $\Omega$ in the ratio $a_0/\overline{\gamma}$, since in the definition of $\Omega$ it is $\sqrt{2}\pi a_0w_0/\overline{\gamma}\lambda \geq 1$. Hence, we could only derive analytical scaling relations in the limit $a_0\to \infty$, which on the other hand would be in conflict with the initial assumption of $\overline{\gamma} \gg a_0$, whence we will not consider this limit in this study.
\section{\label{sec:numerics}Numerical benchmarks}
In the following we are going to test the analytical predictions of Eqs.~(\ref{eq:FD_scatteringangle},\ref{eq:FD_maxscatteringangle},\ref{eq:AD_scatteringangle},\ref{eq:AD_maxscatteringangle}) by numerical simulations. First, we recall that we modeled the transverse laser profile by a step function. As stated above, this assumption naturally overestimates the laser's energy density inside the focal volume by at least a factor $a_0^\textnormal{eff}/a_0 \lesssim 0.7$, where $a_0^\textnormal{eff}$ is the effective laser amplitude. Hence, the resulting constant transverse expulsion of electrons from the laser focus will result in an overestimation of the laser's amplitude, when analyzing the scattered electron bunch's transverse profile. On the other hand, such an overestimation of the laser amplitude is unlikely to change qualitative features of the transverse scattering, but can be modeled as an effective reduction of the parameter $a_0$. In order to provide an estimate for the effective reduction, we first study purely ponderomotive scattering of a randomly distributed initial electron bunch from a Gaussian laser focus with fixed parameters according to the full eq. (\ref{eq:radialmomentum}). The electron bunch is modeled by $10^4$ numerical particles, initially located at a longitudinal position $x_{\|,0} = t_0$ where $t_0<0$ is the start time of the simulation running until $t=-t_0$. The bunch is initially located far away from the laser's focal plane $|x_{\|,0}|\gg l_R$. In transverse direction the electrons will be modeled as randomly distributed in a disk. Finally, we will always consider a laser photon energy of $\omega=1.55$ eV, corresponding to a wavelength of $\lambda = 800$ nm.

\begin{figure}[t]
	\centering\includegraphics[width=.5\linewidth]{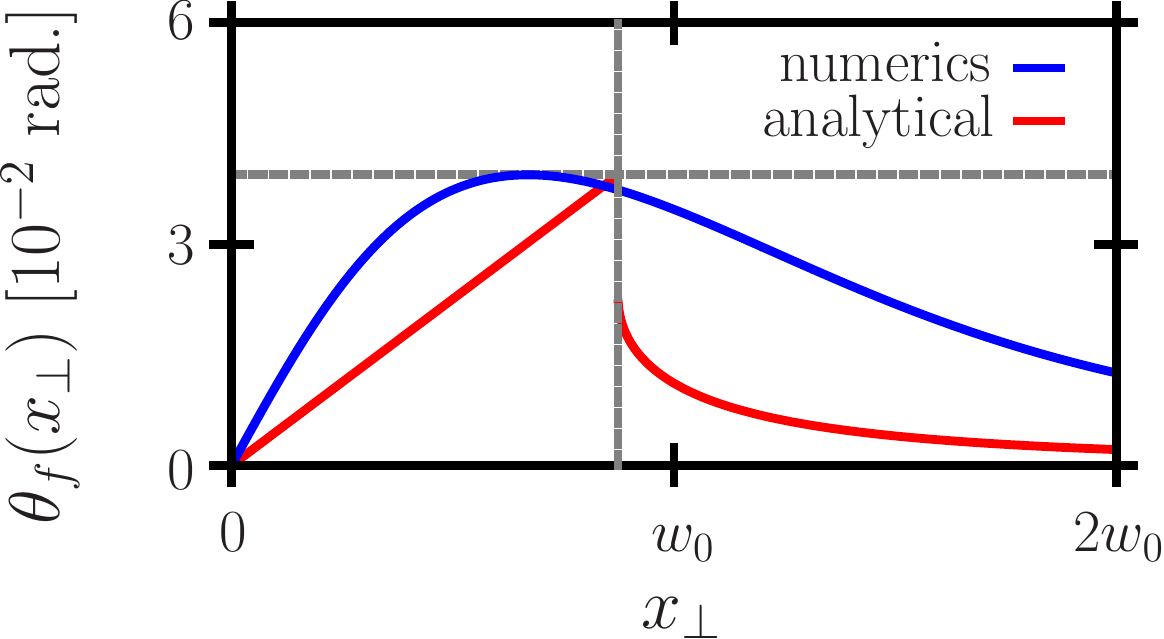}
	\caption{Scattering angle as function of the electron's initial transverse displacement $x_{\perp,0}$ in the focus dominated regime. Since $\Omega>1/2$ in this case, the strongest scattering is experienced by electrons with initial transverse displacement $x_{\perp,0}^\textnormal{peak}$ (dashed vertical line), close to the peak position obtained numerically. The peak scattering angle is accordingly well reproduced by eq. (\ref{eq:FD_scatteringangle}) (dashed horizontal line).}
	\label{fig:Figure2}
\end{figure}
We begin by studying a test case in the focus dominated regime, choosing $a_0=10$, $\varepsilon_i = 200 m$ and $w_0=2.5\, \mu$m, resulting in $\sqrt{2}a_0l_R/(\overline{\gamma}w_0) \approx 0.7$. We will furthermore study an electron disk of initial size $x_{\perp,0} \leq 10 w_0$, in order to account for scattering of particles that will both stay within and leave the laser's focal volume. First, we study the analytically predicted scattering angles as a function of initial transverse displacement. In fact, we find our above conjecture confirmed that in the focus dominated regime those electrons are into largest angles that are initially at a transverse displacement $x_{\perp,0}^\textnormal{peak} = w_0/v(\Omega)$ (s. fig. \ref{fig:Figure2}).

\begin{figure}[b]
	\centering
	\includegraphics[width=\linewidth]{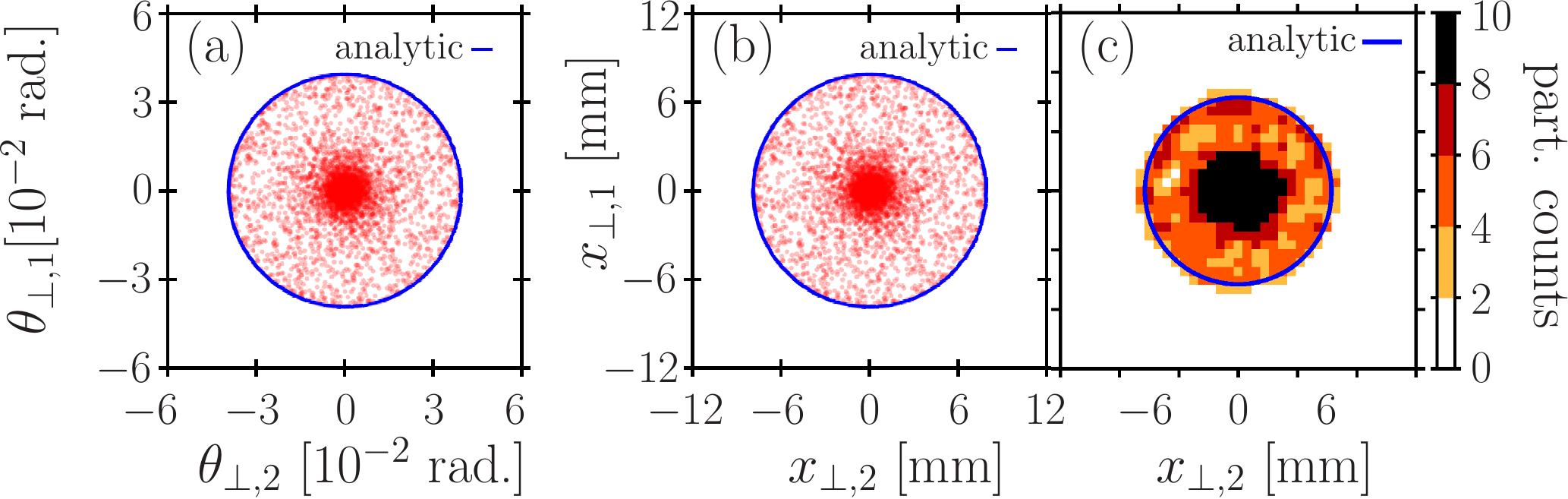}
	\caption{Scattered electron distribution in the focus dominated regime. Every red dot represents a simulated electron's propagation state after interaction with the laser focus. (a): radial velocity distribution after ponderomotive scattering. (b): radial electron positions on a detector screen placed $20$ cm behind the laser focus. (c): same as (b) but with electron data binned at $32\times32$ resolution for $10^4$ electrons in the simulated bunch.}
	\label{fig:Figure1}	
\end{figure}
Next, from a full numerical propagation of the electron trajectories according to Eq.~(\ref{eq:radialmomentum}) we find that at asymptotic times after the interaction with the laser focus the distribution of transverse velocities, which is equivalent to the distribution of scattering angles, is indeed confined to a circular disk (s. fig. \ref{fig:Figure1} (a)). The size of that disk, however, is smaller than the theoretical prediction of eq. (\ref{eq:FD_scatteringangle}), which we attribute to an effective overestimation of $a_0$, as discussed above.
\begin{figure}[t]
	\centering
	\includegraphics[width=.5\linewidth]{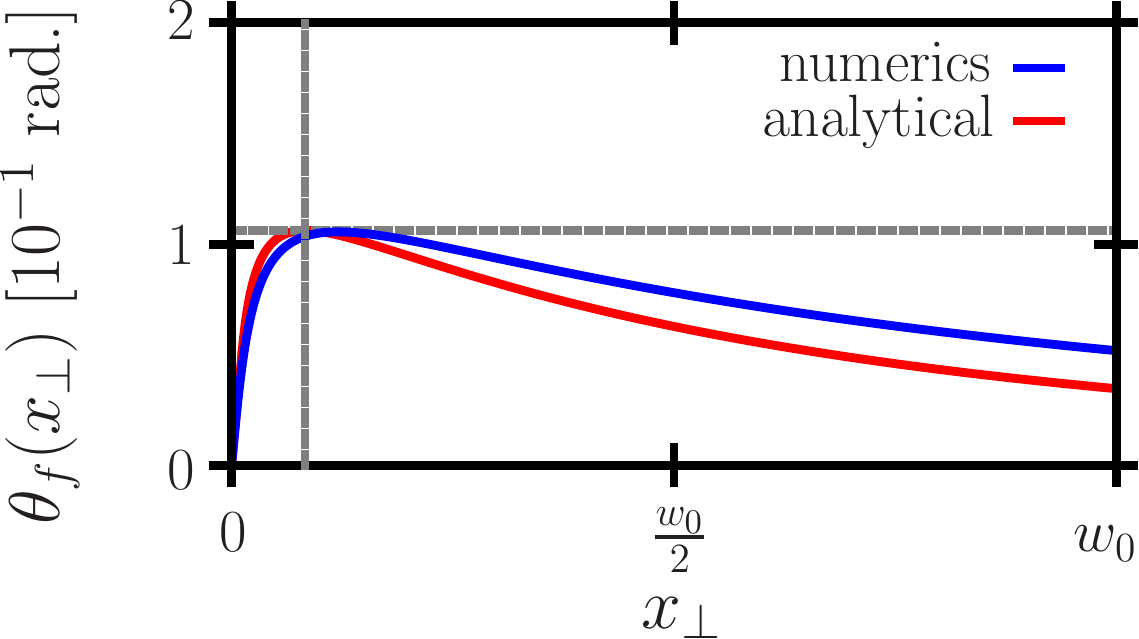}
	\caption{Scattering angle as function of the electron's initial transverse displacement $x_{\perp,0}$ in the amplitude dominated regime. The strongest scattering is experienced by electrons with initial transverse displacement $x_{\perp,0}^\textnormal{peak} = w_0/v(\Omega)>w_0\Omega/\sinh[\pi\Omega]$ (dashed vertical line), close to the peak position obtained numerically. The peak scattering angle is accordingly well reproduced by eq. (\ref{eq:AD_maxscatteringangle}) (dashed horizontal line).}
	\label{fig:Figure8}
\end{figure}
To correct for this effect, we fitted the numerically obtained maximal scattering angle to the theoretical prediction of eq. (\ref{eq:FD_scatteringangle}) with a prefactor of $a_0$ as free parameter. From this procedure, we obtained an effective laser amplitude $a_0^\textnormal{eff}\approx 0.7 a_0$, which upon inserting into eq. (\ref{eq:FD_scatteringangle}) yields the numerically obtained value $\theta_f^\textnormal{max}\approx 3.9\times 10^{-2}$ rad. In order to simulate the signal on an electron detector, we assume the detector to be placed $20$ cm from the laser focus, resulting in a scattered electron bunch radius of approximately $8$ mm (s. fig. \ref{fig:Figure1} (b)). Interestingly, the analytical prediction of the maximal scattering angle $\theta_f^\textnormal{max}$ yields a good agreement with the electrons' spatial distribution, indicating that the assumed detector distance of $20$ cm is sufficient to have the electrons reach their asymptotic propagation state. Finally, in order to emulate the real signal of a pixelated detector, we binned the electrons into a $32\times32$ array, corresponding, e.g., to CCD pixels of $0.75\times0.75$ mm$^2$ size, and find that discarding the saturated detector spot in the bunch center there is still a clear bunch boundary discriminable at the analytically predicted cutoff angle (s. fig. \ref{fig:Figure1} (c)).
\begin{figure}[t]
	\centering
	\includegraphics[width=\linewidth]{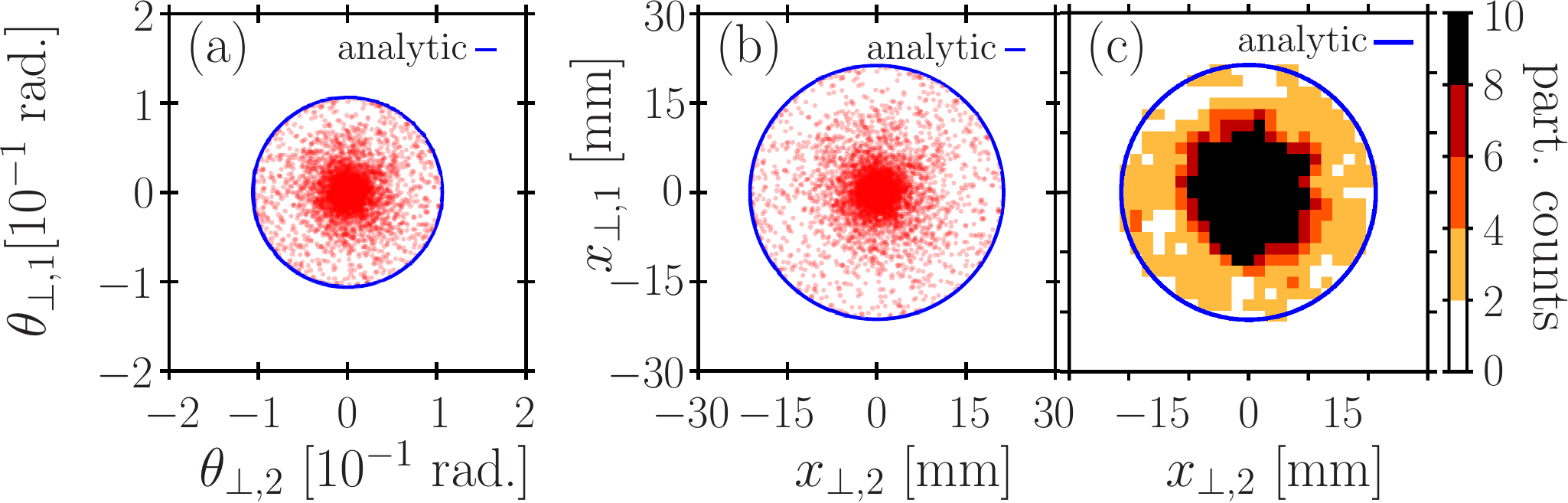}
	\caption{Scattered electron distribution in the amplitude dominated regime. Every red dot represents a simulated electron's propagation state after interaction with the laser focus. Scattering angle as function of the electron's initial transverse displacement $x_{\perp,0}$. The maximum is clearly visible at $x_{\perp,0}=w_0$}
	\label{fig:Figure3}
\end{figure}
%

Next, we study an exemplary case in the amplitude dominated regime. Specifically, we choose $a_0=15$, $\gamma=200$ and $w_0=5\, \mu$m, resulting in $\sqrt{2}a_0l_R/w_0\overline{\gamma} \approx 2.1$. We will again consider the electrons to be initially distributed in a disk of initial size $x_{\perp,0} = 10 w_0$. Studying again the analytically predicted scattering angles as a function of initial transverse displacement, we find that in the amplitude dominated regime, as well, those electrons are scattered into largest angles that are initially at a transverse displacement $x_{\perp,0}^\textnormal{peak} = w_0/v(\Omega)$ (s. fig. \ref{fig:Figure8}). More importantly, we note that even the shape of the numerically simulated electron's scattering angle distribution agrees very well with the analytically derived form, further corroborating our above conjecture that the difference between the model and an experiment can be modeled by an effectively reduced laser peak amplitude.

Also, in analogy to the above discussion, we find that at asymptotic times the transverse electron distribution is confined to a circular disk of opening angle $\theta_f^\textnormal{max}\approx0.11$ rad (s. fig. \ref{fig:Figure3} (a)), larger than the maximal angle obtainable from eq. (\ref{eq:AD_scatteringangle}). Hence, we conclude that as conjectured above, even in the amplitude dominated regime the largest angle scattering is experienced by electrons initially with an initial transverse displacement $x_{\perp,0}^\textnormal{peak} > w_0\Omega/\sinh[\pi \Omega]$. Inserting the corresponding value $x_{\perp,0}^\textnormal{peak}=w_0/v(\Omega)$ into (\ref{eq:AD_maxscatteringangle}) we obtain $\tan \theta (t_\textnormal{max}) \approx 0.11$ in exceptional agreement with the simulated value (s. fig. \ref{fig:Figure3} (a)). Hence, in the amplitude dominated regime we do not need to consider an effectively reduced peak amplitude. The reason for this is most probably that due to the laser's large amplitude the electrons are deflected already long before they reach the laser's focal plane. Far from the laser's focus, however, its transverse Gaussian shape is less significant, rendering the above approximation of it being a step-function more realistic. Assuming again a detector placed at a distance of $20$ cm from the laser focus, we observe the scattered electron bunch radius to be approximately $7$ mm (s. fig. \ref{fig:Figure3} (b)). Again, the agreement between this spatial distribution and the analytical prediction of the maximal scattering angle indicates that at the assumed detector distance the electrons have reached their asymptotic propagation state. Binning the electron distribution again onto a $32\times32$ array we can reconfirm the existence of a clear bunch boundary discriminable at the analytically predicted cutoff angle (s. fig. \ref{fig:Figure3} (c)).


\begin{figure}[t]
	\centering
	\includegraphics[width=.75\linewidth]{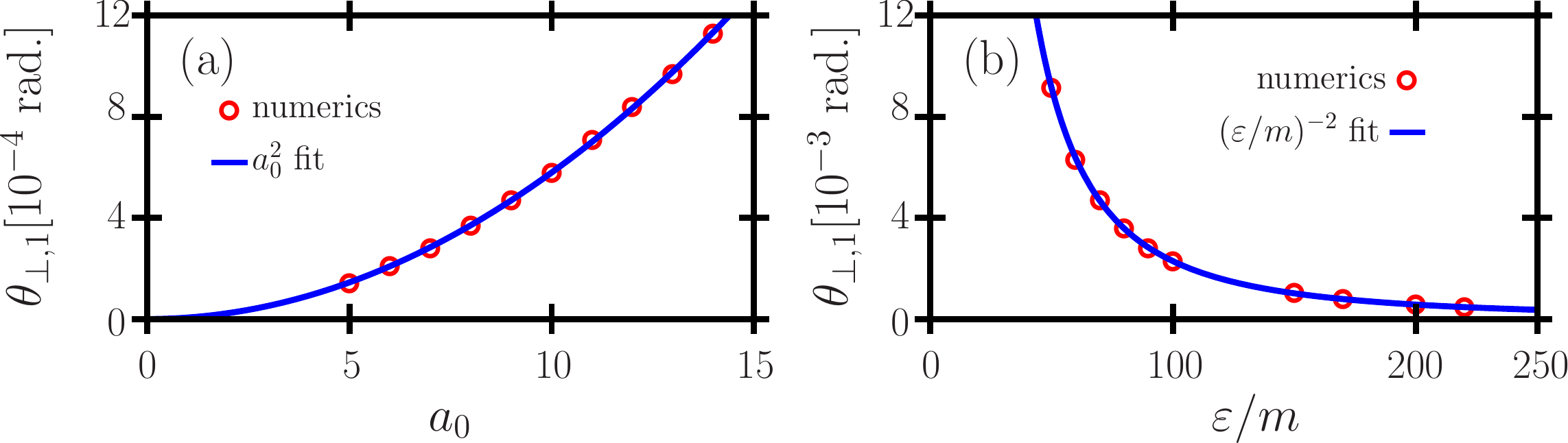}
	\caption{(a) Maximal scattering angle as a function of the laser amplitude $a_0$. (b) Maximal scattering angle as a function of the electron energy $\varepsilon$ (other parameters in the text).}
	\label{fig:Figure4}
\end{figure}

\begin{figure}[b]
 \centering
 \includegraphics[width=.5\linewidth]{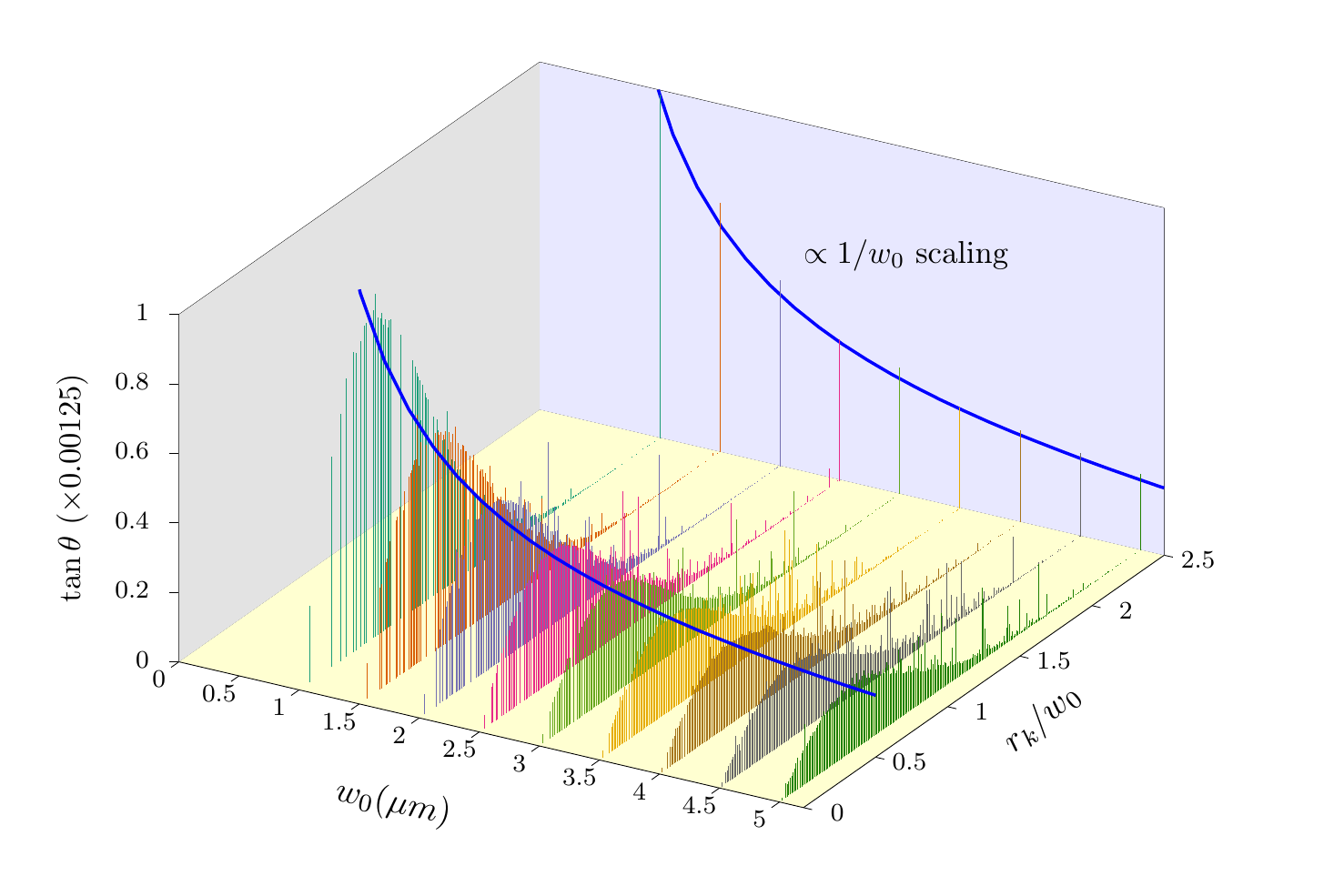}
 \caption{Scattering angles as function of initial transverse displacement $r_k$ and laser focus spot size $w_0$. As a guide to the eye a $1/w_0$ scaling curve (blue) is placed at the position of strongest electron deflection at $r_k = w_0/2$.}
 \label{fig:rk_histo}
\end{figure}
Naturally, it is imperative to test our predictions also beyond the ponderomotive approximation. To this end, we are now going to compare the analytical predictions of the purely ponderomotive effect to a fully numerical simulation of the electron dynamics in the full electrodynamic fields of a Gaussian focused laser. To model such a laser field we use a non-paraxial field approximation \cite{Salamin_Mocken_Keitel_2002}. We simulate an electron bunch of $10^3$ particles, initially placed at a longitudinal position $x_{\|,0} = 500 \lambda$. We begin by testing the predicted scaling laws of the scattering angle. We are again going to study the maximal scattering angles of electrons inside the bunch, which, as demonstrated above, determine the spatial size of the electron bunch's image on a detector, of an electron bunch scattered by the full electromagnetic laser fields of a Gaussian focused laser pulse as functions of different interaction parameters. Studying the maximal scattering angle of an electron with initial energy $\varepsilon = 200 m$ from a laser pulse with focal spot size $w_0 = 2\, \mu$m as a function of the laser amplitude $a_0$, we observe clear reproductions of the above derived scaling laws $\theta_f^\textnormal{max} \sim a_0^2$ (s. fig. \ref{fig:Figure4} (a)). Studying the same scattering for a laser pulse with $a_0=10$ and the same focal spot size $w_0=2\, \mu$m as a function of initial electron energy $\varepsilon$ we find the predicted $\theta_f^\textnormal{max} \sim \varepsilon^{-2}$ scaling well reproduced (s. fig. \ref{fig:Figure4} (b)). In fact, from a fit to the numerical data, we obtain scalings $\theta_f \sim 0.33 (a_0/\gamma)^2$ (solid lines in figs. \ref{fig:Figure4}), in reasonable agreement with the above derived analytical prediction of eqs. (\ref{eq:FD_scatteringangle},\ref{eq:AD_scatteringangle}), corrected for the reduced laser amplitude.

Next to these scaling laws, we also study the maximal scattering angles as functions of the electrons' initial transverse displacement and laser spot sizes (s. fig. \ref{fig:rk_histo}). We find the analytically motivated conjecture that electrons at $x_{\perp,0}\approx w_0/2$ experience the strongest scattering confirmed over a wide range of laser spot sizes. Furthermore, we observe that the for increasing laser spot sizes the maximal deflection angle decreases according to $\theta_f \sim 1/w_0$. While, as argued above, this scaling cannot be derived from our simplified analytical model, it is easily explained by the observation that according to eq. (\ref{eq:radialmomentum}), depending on the electrons' transverse position, they experience radial forces of order $dp_\perp /dt \sim (x_\perp/w_0^2) \exp[-2 (x_\perp/w(x_\|)^2)]$. This force is peaked at $x_\perp \approx 0.71 w(x_\|)$. close to its focal plane, where $w(x_\|)\approx w_0$, but still push electrons to larger $x_\perp$ before they reach this plane. Hence, we expect electrons to experience the strongest radial deflection which start at $x_{\perp,0} \lesssim 0.71 w_0$. As a consequence, we see that the maximal transverse force close to the focal plane scales as $dp_\perp /dt \sim 1/w_0$.

\section{\label{sec:detection}Experimental realisation}

For an experimental implementation of ponderomotive electron scattering for a reliable $a_0$-measurement, the interaction must take place in the amplitude-dominated regime. Otherwise, the scattered electrons would not leave the focusing cone of the laser beam and hit -- due to the head-on geometry -- the focusing optic of the laser. In the following we will show that for any realistic experimental use, this implies a maximum initial electron energy, determined by the laser power and, as a minor correction, by the focusing strength that can be used. We will also argue towards an optimum electron energy, such that the electrons' scattering angles are sufficiently large for them to significantly separate from the laser cone, facilitating a safe measurement of their scattered profile, reasonably unaffected by the laser's optics.

Assuming the unfocussed laser beam to have a Gaussian profile with waist $w_{\rm{beam}}$, the beam size in the focus $w_0$ is obtained via optics with a focal length $f$ given by $w_0 = \lambda f/(\pi w_{\rm{beam}})$. It can be shown that $d \approx 3 w_{\rm{beam}}$ is a useful diameter of the focusing optics (or effective beam diameter). Then, the intensity on the optics' boundary is about 1\% of the peak intensity and the laser energy loss due to clipping is also about 1\%, which can be considered as negligible. Consequently, from the F-number of the laser focusing $F_{\#} \equiv f/d $ we derive the focus spot size to be given by $w_0 = 3 \lambda F_{\#} / \pi$.

Next, the experimentalist's expression of $a_0$, derived from the laser intensity $I$, is 
\begin{equation}     \label{eq:EXP_a0} 
a_0 = 0.85 \frac{\lambda}{\rm{\mu m}} \sqrt{\frac{I}{10^{18} \rm{W/cm^2}}}\;.
\end{equation} 
Using $I=P/(2\pi w_0^2)$ with the laser power $P$ and the above expression for $w_0$, one can find 
\begin{equation} \label{eq:EXP_a0^2}
    a_0^2 = 100\left(\frac{0.85}{3 F_{\#}}\right)^2 \frac{\pi}{2} \frac{P}{\rm{TW}} \;.
\end{equation} 
We now recall from Eq.~(\ref{eq:xieom}) that the amplitude-dominated case requires $   \sqrt{2}a_0/\overline{\gamma} > w_0/l_R$ where the Lorentz factor $\overline{\gamma}$ was assumed to be constant and to decompose as $\overline{\gamma}^2 = \gamma_0^2 + a_0^2$ with $\gamma_0 = \sqrt{1+(\varepsilon/m)^2}$ the initial Lorentz factor before the scattering. Now, this condition for the parameter set to lie in the amplitude dominated regime can be solved for $\gamma_0$ with the previous relations to
\begin{eqnarray} \label{eq:EXP_solution-gamma_0-full}
    \gamma_0^2 < (\gamma_{0}^{\rm{max}})^2 = 100 \pi (0.85)^2 \frac{P}{\rm{TW}} \left(1- \frac{1}{2(3 F_{\#})^2} \right)\;,
\end{eqnarray} 
setting an upper limit for the initial electron energy.

Next, we recall that for the derivation of the scattering angles, see Sec.~\ref{sec:analytics}, the constraints were both $\gamma_0 \gg a_0$ and $\gamma_0 \gg 1$, forming a set of lower limits. For typical focusing systems with $2 < F_{\#} < 20$, the correction due to $F_{\#}$ is of percent-level or below and can be neglected. Hence the maximum initial Lorentz factor can be approximated to 
\begin{equation} \label{eq:EXP_approx-gamma_0}
    \gamma_{0}^{\rm{max}} \approx 15 \sqrt{\frac{P}{\rm{TW}}}
\end{equation} 
which, for relevant multi-TW to PW laser systems, is of the order of $ \gamma_{0}^{\rm{max}} \approx 50 \dots 500$. Hence, $\gamma_0 \gg 1$ can be easily fulfilled without violating Eq.~(\ref{eq:EXP_solution-gamma_0-full}). Only for weak laser powers of TW-level or below one would transit from the amplitude-dominated regime to the focus-dominated regime with too shallow scattering angles for detection.

On the other hand, also $\gamma_0 \gg a_0$ must hold for the discussed scattering process (regardless the regime), effectively setting the lower limit for $\gamma_0$. Eqns. (\ref{eq:EXP_a0^2}, \ref{eq:EXP_solution-gamma_0-full}) show that $\gamma_{0}^{\rm{max}} \approx 3\sqrt{2} F_{\#} a_0$. As result, the window for the initial electron energy is given by
\begin{equation} \label{eq:EXP_gamma_0_window}
    a_0^2 \ll \gamma_{0}^2 < 18 F_{\#}^2 a_0^2 \;.
\end{equation} 
The longer the F-number, the larger the range for $\gamma_{0}$ is: For relatively tight focusing with $F_{\#} = 2.35$, there is just a factor 10 range for $\gamma_0$ to be between $\gamma_0 = a_0$ and $\gamma_0 = \gamma_{0}^{\rm{max}}$. For $F_{\#} = 10$, the range is already a factor 40 and both conditions $\gamma_0 \gg a_0$ and  $\gamma_0 < \gamma_{0}^{\rm{max}}$ can be easily fulfilled.

We stress, however, that despite this obvious necessity to optimise the interaction parameters, the above studied case used $a_0=15$ and for a loose focusing of $F_\#=10$ already satisfied $\gamma_0=200\ll \gamma_0^{\rm{max}} \approx 636$. The resulting scattering angles were found to be $\theta_f^\textnormal{peak}\approx 0.1> \arctan[2/(3 F_{\#})] \approx 0.06$, where the arctan expression assumes the optic's mount to add $1/3$ of the optic's size to the blocked solid angle. This indicates that the above studied numerical example carries the potential of experimental observation with the electron scattered into angles large enough for them to decouple from required focusing optics.

\section{\label{sec:conclusion}Conclusion}
We have derived a novel solution of the ponderomotive scattering of ultra-relativistic electrons off intense focused laser pulses. We demonstrated and verified numerically that an electron's scattering angle is determined by a simple ratio of the laser's field amplitude to the electron's initial energy. Hence, we could provide simple scaling laws that may possibly allow to read off a focused laser's intensity from the spatial scattering distribution of an externally accelerated electron bunch brought into collision with the laser pulse.

\section*{\label{sec:acknowledgements}Acknowledgements}
AH acknowledges the Science and Engineering Research Board, Department of Science and Technology, Government of India for funding the project EMR/2016/002675. AH also acknowledges the local hospitality and the travel support of the Max Planck Institute for the Physics of Complex Systems, Dresden, Germany.


\vskip 5mm
\begin{thebibliography}{10}

\bibitem{VulcanLaser}
C.N. Danson, P.A. Brummitt, R.J. Clarke, J.L. Collier, B.~Fell, A.J.
  Frackiewicz, S.~Hancock, S.~Hawkes, C.~Hernandez-Gomez, P.~Holligan, M.H.R.
  Hutchinson, A.~Kidd, W.J. Lester, I.O. Musgrave, D.~Neely, D.R. Neville, P.A.
  Norreys, D.A. Pepler, C.J. Reason, W.~Shaikh, T.B. Winstone, R.W.W. Wyatt,
  and B.E. Wyborn.
\newblock Vulcan petawatt—an ultra-high-intensity interaction facility.
\newblock {\em Nuclear Fusion}, 44(12):S239, 2004.

\bibitem{Mourou_etal_2006}
G.A. Mourou, T.~Tajima, and S.V. Bulanov.
\newblock Optics in the relativistic regime.
\newblock {\em Rev. Mod. Phys.}, 78:309--371, Apr 2006.

\bibitem{Hooker_etal_2008}
C.J. Hooker, S.~Blake, O.~Chekhlov, R.J. Clarke, J.L. Collier, E.J. Divall,
  K.~Ertel, P.S. Foster, S.J. Hawkes, P.~Holligan, B.~Landowski, B.J. Lester,
  D.~Neely, B.~Parry, R.~Pattathil, M.~Streeter, and B.E. Wyborn.
\newblock Commissioning the astra gemini petawatt ti:sapphire laser system.
\newblock In {\em Conference on Lasers and Electro-Optics/Quantum Electronics
  and Laser Science Conference and Photonic Applications Systems Technologies},
  page JThB2. Optical Society of America, 2008.

\bibitem{Leemans_etal_2010}
W.~P. Leemans, R.~Duarte, E.~Esarey, S.~Fournier, C.~G.~R. Geddes, D.~Lockhart,
  C.~B. Schroeder, C.~Toth, J.‐L. Vay, and S.~Zimmermann.
\newblock The berkeley lab laser accelerator (bella): A 10 gev laser plasma
  accelerator.
\newblock {\em AIP Conference Proceedings}, 1299(1):3--11, 2010.

\bibitem{Zou_etal_2015}
J.P. Zou, C.~Le~Blanc, D.N. Papadopoulos, G.~Chériaux, P.~Georges,
  G.~Mennerat, F.~Druon, L.~Lecherbourg, A.~Pellegrina, P.~Ramirez, and et~al.
\newblock Design and current progress of the apollon 10 pw project.
\newblock {\em High Power Laser Sci. Eng.}, 3:e2, 2015.

\bibitem{Kawanaka_etal_2016}
J.~Kawanaka, K.~Tsubakimoto, H.~Yoshida, K.~Fujioka, Y.~Fujimoto, S.~Tokita,
  T.~Jitsuno, N.~Miyanaga, and Gekko-EXA~Design Team.
\newblock Conceptual design of sub-exa-watt system by using optical parametric
  chirped pulse amplification.
\newblock {\em Journal of Physics: Conference Series}, 688(1):012044, 2016.

\bibitem{Gales_etal_2018}
S~Gales, K~A Tanaka, D~L Balabanski, F~Negoita, D~Stutman, O~Tesileanu, C~A Ur,
  D~Ursescu, I~Andrei, S~Ataman, M~O Cernaianu, L~D’Alessi, I~Dancus,
  B~Diaconescu, N~Djourelov, D~Filipescu, P~Ghenuche, D~G Ghita, C~Matei,
  K~Seto, M~Zeng, and N~V Zamfir.
\newblock The extreme light infrastructure—nuclear physics (eli-np) facility:
  new horizons in physics with 10 pw ultra-intense lasers and 20 mev brilliant
  gamma beams.
\newblock {\em Reports on Progress in Physics}, 81(9):094301, 2018.

\bibitem{CORELS}
{CORELS
  \href{https://www.ibs.re.kr/eng/sub02_03_05.do}{https://www.ibs.re.kr/eng/sub02\_03\_05.do}}.

\bibitem{Draco}
U.~Schramm, M.~Bussmann, A.~Irman, M.~Siebold, K.~Zeil, D.~Albach, C.~Bernert,
  S.~Bock, F.~Brack, J.~Branco, J.~P. Couperus, T.~E. Cowan, A.~Debus,
  C.~Eisenmann, M.~Garten, R.~Gebhardt, S.~Grams, U.~Helbig, A.~Huebl,
  T.~Kluge, A.~Koehler, J.~M. Kraemer, S.~Kraft, F.~Kroll, M.~Kuntzsch,
  U.~Lehnert, M.~Loeser, J.~Metzkes, P.~Michel, L.~Obst, R.~Pausch, M.~Rehwald,
  R.~Sauerbrey, H.~P. Schlenvoigt, K.~Steiniger, and O.~Zarini.
\newblock {First results with the novel petawatt laser acceleration facility in
  Dresden}.
\newblock In {\em {8TH INTERNATIONAL PARTICLE ACCELERATOR CONFERENCE (IPAC
  2017)}}, volume {874} of {\em {Journal of Physics Conference Series}}.
  {European Spallat Source; European Phys Soc Accelerator Grp; Int Union Pure
  \& Appl Phys}, {2017}.
\newblock {8th International Particle Accelerator Conference (IPAC), Bella Ctr,
  Copenhagen, DENMARK, MAY 14-19, 2017}.

\bibitem{DiPiazza_etal_2012}
A.~{Di Piazza}, C.~Mueller, K.~Z. Hatsagortsyan, and C.~H. Keitel.
\newblock Extremely high-intensity laser interactions with fundamental quantum
  systems.
\newblock {\em Rev. Mod. Phys.}, 84(3):1177--1228, AUG 16 2012.

\bibitem{Sarachik_Schappert_1970}
E.~S. Sarachik and G.~T. Schappert.
\newblock Classical theory of the scattering of intense laser radiation by free
  electrons.
\newblock {\em Phys. Rev. D}, 1:2738--2753, May 1970.

\bibitem{Salamin_Faisal_1996}
Y.I. Salamin and F.H.M. Faisal.
\newblock Harmonic generation by superintense light scattering from
  relativistic electrons.
\newblock {\em Phys. Rev. A}, 54:4383--4395, Nov 1996.

\bibitem{Woodward_1946}
P.~M. {Woodward}.
\newblock A method of calculating the field over a plane aperture required to
  produce a given polar diagram.
\newblock {\em Journal of the Institution of Electrical Engineers - Part IIIA:
  Radiolocation}, 93(10):1554--1558, 1946.

\bibitem{Lawson_1979}
J.~D. {Lawson}.
\newblock Lasers and accelerators.
\newblock {\em IEEE Transactions on Nuclear Science}, 26(3):4217--4219, June
  1979.

\bibitem{LandauI}
L.D. Landau and E.~M. Lifshitz.
\newblock {\em Mechanics}.
\newblock Elsevier Butterworth-Heinemann, Oxford, 1976.

\bibitem{Schmidt_1979}
In George Schmidt, editor, {\em Physics of High Temperature Plasmas}. Academic
  Press, second edition, 1979.

\bibitem{Boot_etal_1958}
H.A.H. Boot, S.A. Self, and R.B. R-Shersby-Harvie.
\newblock Containment of a fully-ionized plasma by radio-frequency fields.
\newblock {\em Journal of Electronics and Control}, 4(5):434--453, 1958.

\bibitem{Gaponov_Miller_1958}
A.V. Gaponov and M.A. Miller.
\newblock Potential wells for charged particles in a high-frequency
  electromagnetic field.
\newblock {\em Sov. Phys. JETP}, 7:168, 1958.

\bibitem{Kibble_1966}
T.~W.~B. Kibble.
\newblock Mutual refraction of electrons and photons.
\newblock {\em Phys. Rev.}, 150:1060--1069, Oct 1966.

\bibitem{Hopf_etal_1976}
F.~A. Hopf, P.~Meystre, M.~O. Scully, and W.~H. Louisell.
\newblock Strong-signal theory of a free-electron laser.
\newblock {\em Phys. Rev. Lett.}, 37:1342--1345, Nov 1976.

\bibitem{Bauer_etal_1995}
D.~Bauer, P.~Mulser, and W.~H. Steeb.
\newblock Relativistic ponderomotive force, uphill acceleration, and transition
  to chaos.
\newblock {\em Phys. Rev. Lett.}, 75:4622--4625, Dec 1995.

\bibitem{Hartemann_etal_1995}
F.~V. Hartemann, S.~N. Fochs, G.~P. Le~Sage, N.~C. Luhmann, J.~G. Woodworth,
  M.~D. Perry, Y.~J. Chen, and A.~K. Kerman.
\newblock Nonlinear ponderomotive scattering of relativistic electrons by an
  intense laser field at focus.
\newblock {\em Phys. Rev. E}, 51:4833--4843, May 1995.

\bibitem{Mora_Antonsen_1997}
P.~Mora and T.M. Antonsen, Jr.
\newblock Kinetic modeling of intense, short laser pulses propagating in
  tenuous plasmas.
\newblock {\em Physics of Plasmas}, 4(1):217--229, 1997.

\bibitem{Salamin_Faisal_1997}
Y.I. Salamin and F.H.M. Faisal.
\newblock Ponderomotive scattering of electrons in intense laser fields.
\newblock {\em Phys. Rev. A}, 55:3678--3683, May 1997.

\bibitem{Hartemann_etal_1998}
F.~V. Hartemann, J.~R. Van~Meter, A.~L. Troha, E.~C. Landahl, N.~C. Luhmann,
  H.~A. Baldis, Atul Gupta, and A.~K. Kerman.
\newblock Three-dimensional relativistic electron scattering in an
  ultrahigh-intensity laser focus.
\newblock {\em Phys. Rev. E}, 58:5001--5012, Oct 1998.

\bibitem{Quesnel_Mora_1998}
B.~Quesnel and P.~Mora.
\newblock Theory and simulation of the interaction of ultraintense laser pulses
  with electrons in vacuum.
\newblock {\em Phys. Rev. E}, 58:3719--3732, Sep 1998.

\bibitem{Bituk_Fedorov_1999}
M.~V. Bituk, D. R.and~Fedorov.
\newblock Relativistic ponderomotive forces.
\newblock {\em Journal of Experimental and Theoretical Physics},
  89(4):640--646, Oct 1999.

\bibitem{Narozhny_Fofanov_2000}
N.~B. Narozhny and M.~S. Fofanov.
\newblock Scattering of relativistic electrons by a focused laser pulse.
\newblock {\em Journal of Experimental and Theoretical Physics},
  90(5):753--768, May 2000.

\bibitem{Yu_etal_2000b}
W.~Yu, M.~Y. Yu, J.~X. Ma, Z.~M. Sheng, J.~Zhang, H.~Daido, S.~B. Liu, Z.~Z.
  Xu, and R.~X. Li.
\newblock Ponderomotive acceleration of electrons at the focus of high
  intensity lasers.
\newblock {\em Phys. Rev. E}, 61:R2220--R2223, Mar 2000.

\bibitem{Castillo_Milantev_2014}
A.~J. Castillo and V.~P. Milant'ev.
\newblock Relativistic ponderomotive forces in the field of intense laser
  radiation.
\newblock {\em Technical Physics}, 59(9):1261--1266, Sep 2014.

\bibitem{Shiryaev_2019}
O.B. Shiryaev.
\newblock Asymptotic theory of the ponderomotive dynamics of an electron driven
  by a relativistically intense focused electromagnetic envelope.
\newblock {\em arXiv:1901.02335v2}, 2019.

\bibitem{Moore_etal_1995}
C.~I. Moore, J.~P. Knauer, and D.~D. Meyerhofer.
\newblock Observation of the transition from thomson to compton scattering in
  multiphoton interactions with low-energy electrons.
\newblock {\em Phys. Rev. Lett.}, 74:2439--2442, Mar 1995.

\bibitem{Malka_etal_1997}
G.~Malka, E.~Lefebvre, and J.~L. Miquel.
\newblock Experimental observation of electrons accelerated in vacuum to
  relativistic energies by a high-intensity laser.
\newblock {\em Phys. Rev. Lett.}, 78:3314--3317, Apr 1997.

\bibitem{Salamin_Mocken_Keitel_2002}
Y.I. Salamin, G.R. Mocken, and C.H. Keitel.
\newblock Electron scattering and acceleration by a tightly focused laser beam.
\newblock {\em Phys. Rev. ST Accel. Beams}, 5:101301, Oct 2002.

\bibitem{Liu_etal_2008}
Y.~Liu, J.~Zhang, H.~Wu, and Z.~Sheng.
\newblock Ponderomotive scattering of electrons and its application to measure
  the pulse duration of ultrafast electron beams.
\newblock {\em Journal of Applied Physics}, 103(4):044905, 2008.

\bibitem{Zhu_etal_2018}
P.~Zhu, X.~Xie, J.~Kang, Q.~Yang, H.~Zhu, A.~Guo, M.~Sun, Q.~Gao, Z.~Cui,
  X.~Liang, S.~Yang, D.~Zhang, and J.~Zhu.
\newblock {Systematic study of spatiotemporal influences on temporal contrast
  in the focal region in large-aperture broadband ultrashort petawatt lasers}.
\newblock {\em High Power Laser Science and Engineering}, 6:1--7, 2018.

\bibitem{Bor_1988}
Z.~Bor.
\newblock {Distortion of femtosecond laser pulses in lenses and lens systems}.
\newblock {\em Journal of Modern Optics}, 35(12):1907--1918, 1988.

\bibitem{Heuck_etal_2006}
H.-M. Heuck, P.~Neumayer, T.~K{\"u}hl, and U.~Wittrock.
\newblock Chromatic aberration in petawatt-class lasers.
\newblock {\em Applied Physics B}, 84(3):421--428, Sep 2006.

\bibitem{Wu_etal_2018}
Y.~P. Wu, J.~F. Hua, C.~H. Pai, X.~L. Xu, C.~J. Zhang, F.~Li, Y.~Wan, Z.~Nie,
  W.~Lu, W.~B. Mori, and C.~Joshi.
\newblock A near-ideal dechirper for plasma based electron and positron
  acceleration using a hollow channel plasma.
\newblock {\em arXiv:1805.07031}, 2018.

\bibitem{Link_etal_2006}
A.~Link, E.A. Chowdhury, J.T. Morrison, V.M. Ovchinnikov, D.~Offermann,
  L.~Van~Woerkom, R.R. Freeman, J.~Pasley, E.~Shipton, F.~Beg, P.~Rambo,
  J.~Schwarz, M.~Geissel, A.~Edens, and J.L. Porter.
\newblock Development of an in situ peak intensity measurement method for
  ultraintense single shot laser-plasma experiments at the sandia z petawatt
  facility.
\newblock {\em Review of Scientific Instruments}, 77(10):10E723, 2006.

\bibitem{Smeenk_etal_2011}
C.~Smeenk, J.~Z. Salvail, L.~Arissian, P.~B. Corkum, C.~T. Hebeisen, and
  A.~Staudte.
\newblock Precise in-situ measurement of laser pulse intensity using strong
  field ionization.
\newblock {\em Opt. Express}, 19(10):9336--9344, May 2011.

\bibitem{Ciappina_etal_2019}
M.~F. Ciappina, S.~V. Popruzhenko, S.~V. Bulanov, T.~Ditmire, G.~Korn, and
  S.~Weber.
\newblock Progress toward atomic diagnostics of ultrahigh laser intensities.
\newblock {\em Phys. Rev. A}, 99:043405, Apr 2019.

\bibitem{Galkin_etal_2010}
A.~L. Galkin, M.~P. Kalashnikov, V.~K. Klinkov, V.~V. Korobkin, M.~Yu.
  Romanovsky, and O.~B. Shiryaev.
\newblock {Electrodynamics of electron in a superintense laser field: New
  principles of diagnostics of relativistic laser intensity}.
\newblock {\em {PHYSICS OF PLASMAS}}, {17}({5}), {MAY} {2010}.

\bibitem{Kalashnikov_etal_2015}
M.~Kalashnikov, A.~Andreev, K.~Ivanov, A.~Galkin, V.~Korobkin, M.~Romanovsky,
  O.~Shiryaev, M.~Schnuerer, J.~Braenzel, and V.~Trofimov.
\newblock {Diagnostics of peak laser intensity based on the measurement of
  energy of electrons emitted from laser focal region}.
\newblock {\em {LASER AND PARTICLE BEAMS}}, {33}({3}):{361--366}, {SEP} {2015}.

\bibitem{Ivanov_etal_2018}
K.~A. Ivanov, I.~N. Tsymbalov, O.~E. Vais, S.~G. Bochkarev, R.~V. Volkov, V.~Yu
  Bychenkov, and A.~B. Savel'Ev.
\newblock {Accelerated electrons for in situ peak intensity monitoring of
  tightly focused femtosecond laser radiation at high intensities}.
\newblock {\em Plasma Physics and Controlled Fusion}, 2018.

\bibitem{Fuchs2006NJP}
J.~Fuchs, P.~Antici, E.~D'Humi{\`{e}}res, E.~Lefebvre, M.~Borghesi,
  E.~Brambrink, C.~A. Cecchetti, M.~Kaluza, V.~Malka, M.~Manclossi,
  S.~Meyroneinc, P.~Mora, J.~Schreiber, T.~Toncian, H.~P{\'{e}}pin, and
  P.~Audebert.
\newblock {Laser-driven proton scaling laws and new paths towards energy
  increase}.
\newblock {\em Nature Physics}, 2(1):48--54, 2006.

\bibitem{Daido_etal_2012}
H.~Daido, M.~Nishiuchi, and A.S. Pirozhkov.
\newblock Review of laser-driven ion sources and their applications.
\newblock {\em Rep. Prog. Phys.}, 75(5):056401, May 2012.

\bibitem{Macchi_etal_2013}
A.~Macchi, M.~Borghesi, and M.~Passoni.
\newblock Ion acceleration by superintense laser-plasma interaction.
\newblock {\em Rev. Mod. Phys.}, 85:751--793, May 2013.

\bibitem{Mackenroth_etal_2017a}
F.~Mackenroth, A.~Gonoskov, and M.~Marklund.
\newblock Reaching high flux in laser-driven ion acceleration.
\newblock {\em The European Physical Journal D}, 71(8):204, Aug 2017.

\bibitem{Zeil2010NJP}
K.~Zeil, S.~D. Kraft, S.~Bock, M.~Bussmann, T.~E. Cowan, T.~Kluge, J.~Metzkes,
  T.~Richter, R.~Sauerbrey, and U.~Schramm.
\newblock {The scaling of proton energies in ultrashort pulse laser plasma
  acceleration}.
\newblock {\em New Journal of Physics}, 12, 2010.

\bibitem{Mackenroth_etal_2010}
F.~Mackenroth, A.~{Di Piazza}, and C.~H. Keitel.
\newblock Determining the carrier-envelope phase of intense few-cycle laser
  pulses.
\newblock {\em Phys. Rev. Lett.}, 105:063903, Aug 2010.

\bibitem{Har-Shemesh_etal_2012}
O.~Har-Shemesh and A.~{Di Piazza}.
\newblock Peak intensity measurement of relativistic lasers via nonlinear
  thomson scattering.
\newblock {\em Opt. Lett.}, 37(8):1352, Apr 2012.

\bibitem{Li_etal_2018}
Jian-Xing Li, Yue-Yue Chen, Karen~Z. Hatsagortsyan, and Christoph~H. Keitel.
\newblock Single-shot carrier-envelope phase determination of long superintense
  laser pulses.
\newblock {\em Phys. Rev. Lett.}, 120:124803, Mar 2018.

\bibitem{Harvey_etal_2018}
C.~N. Harvey.
\newblock In situ characterization of ultraintense laser pulses.
\newblock {\em Phys. Rev. Accel. Beams}, 21:114001, Nov 2018.

\bibitem{Mackenroth_Holkundkar_2017}
F.~Mackenroth and A.R. Holkundkar.
\newblock Determining the duration of an intense laser pulse directly in focus.
\newblock {\em arXiv:1712.06898}, 2017.

\bibitem{Boca_Florescu_2009}
M.~Boca and V.~Florescu.
\newblock Nonlinear compton scattering with a laser pulse.
\newblock {\em Phys. Rev. A}, 80:053403, Nov 2009.

\bibitem{Mackenroth_DiPiazza_2011}
F.~Mackenroth and A.~{Di Piazza}.
\newblock Nonlinear compton scattering in ultrashort laser pulses.
\newblock {\em Phys. Rev. A}, 83:032106, Mar 2011.

\bibitem{Mackenroth_DiPiazza_2013}
F.~Mackenroth and A.~Di~Piazza.
\newblock Nonlinear double compton scattering in the ultrarelativistic quantum
  regime.
\newblock {\em Phys. Rev. Lett.}, 110:070402, Feb 2013.

\end{thebibliography}
\end{document}